\documentclass[aps,prb,reprint,showpacs]{revtex4-2}
\usepackage[english]{babel}
\usepackage{amsmath,amssymb,bbm,graphicx,color,comment,txfonts}
\usepackage[bookmarks=true,colorlinks,citecolor=blue,linkcolor=blue,urlcolor=blue]{hyperref}
\usepackage{dsfont}
\definecolor{darkgreen}{rgb}{0,0.5,0}
\definecolor{orange}{rgb}{1,0.5,.3}

\usepackage{bbm}
\usepackage{float}

\usepackage{dcolumn}   
\usepackage{bm}        
\usepackage{lineno}

\usepackage{mathtools}
\usepackage{xcolor}
\usepackage{tabularx}
\usepackage{multirow}

\usepackage[export]{adjustbox}
\usepackage{sidecap}

\usepackage{braket}

\newcommand\ii{\mathrm{i}}

\newcommand{\mean}[1]{\langle #1 \rangle}

\newcommand{\cre}[2]{{#1}^{\dagger}_{#2}}
\newcommand{\ann}[2]{{#1}^{\phantom{\dagger}}_{#2}}
\newcommand{\im}[1]{\text{Im } #1}

\newcommand{\code}[1]{\texttt{#1}}

\newcommand\bigO{\mathcal{O}}

\begin{document}

\title{Merging numerical renormalization group and intermediate representation to compactify two- and three-point correlators}
\author{Sebastian Huber}
\author{Markus Wallerberger}
\author{Paul Worm}
\author{Karsten Held}
\date{\today}
\affiliation{Institute of Solid State Physics, TU Wien, 1040 Vienna, Austria}

\begin{abstract}
  The vanguard of many-body theory  is nowadays dealing with the
 full frequency dynamics of  
 $n$-point Green's functions for  $n$ higher than two. Numerically,
 these objects easily become a memory bottleneck, even when
 working with discrete imaginary-time Matsubara frequencies.
 Here, we use the  intermediate representation (IR) to compactify the two-point Green's function and three-point Fermion-Bose vertex directly on the real frequency axis, on the basis of numerical renormalization group (NRG) data.
 We  empirically observe  an upper bound of the relative error when comparing the IR reconstructed signal with the original NRG data, and demonstrate
that a IR compacification is possible.
\end{abstract}

\maketitle

\section{Introduction}\label{Intro}

Many-body $n$-point
Green's functions are  at the very core of quantum field theory \cite{abrikosovmethods,mahanmany}.
Its  two-point variant describes the propagation of a single particle, which may be strongly  renormalized or even indicate a Mott metal-insulator transition due to strong interactions \cite{Gebhard1997,Imada1998}.
The three-point Green's function  describes the coupling of three bosons or that of fermions to bosonic degrees of freedom. An example is the coupling of electrons to spin fluctuations, which, among others, is of relevance for the physics of high-temperature superconductivity \cite{Monthoux1991,Abanov2003,Huang2006,Krien2021b}. The four-point fermionic Green's function, in turn,  is connected to physical response functions. An example is the optical conductivity which may show various vertex corrections, such as excitons \cite{Frenkel1931,Wannier37}, weak-localization corrections \cite{Altshuler1985} and $\pi$-tons \cite{Kauch2020}.

Such $n$-point Green's functions or the associated vertex functions also form the backbone of dynamical mean-field theory (DMFT) \cite{Metzner1989,Georges1992,Georges1996,Held2008} and diagrammatic extensions thereof \cite{Toschi2007,Rubtsov2008,RMPvertex}. Specifically, DMFT is closely connected to the two-point (one-particle) Green's function or the associated vertex, the self-energy, which are determined self-consistently within the DMFT cycle \cite{Georges1992}. Diagrammatic extensions
are either based on the four-point vertex  \cite{Toschi2007,Rubtsov2008}, where the three-point vertex enters additionally for solving the Bethe-Salpeter equation efficiently \cite{Katanin2009,Galler2019}, or on the three-point vertex directly \cite{Ayral2016a,Stepanov2021}. Also  tiling a major part of the non-local correlations with three-point vertices and including the residual four-point vertex in the parquet equations is possible \cite{Krien2020,Krien2021}.

These three-point(and even more so the four-point) correlators easily become huge objects in the full parquet version of the diagrammatic extensions of DMFT \cite{Valli2015,Li2017},  where they depend on two(three) frequencies and momenta. Momentum is discretized within the Brillouin zone, and frequencies are discretized at Matsubara frequencies. Despite  using a truncated-unity expansion for the momenta  \cite{Eckhardt2020} and the  high-frequency asymptotics for the frequencies \cite{Li2017,Wentzell2020}, memory access quickly becomes the computational bottleneck.

A breakthrough to overcome this obstacle was recently achieved using the intermeditate representation (IR) \cite{Shinaoka2017,shinaoka2018overcomplete} which is based on the rapid decay of singular values of the integration kernel for the spectral representation in terms of Matsubara frequencies.
At the same time, pioneering work using the numerical renormalization group (NRG)  \cite{kugler2021multipoint,lee2021computing} allowed, for the first time, the calculation of the full real-frequency dynamics of the four-point correlator for a single impurity  Anderson  model (SIAM). This four-point Green's function or the associated vertex constitute the starting point for the aforementioned diagrammatic extensions of DMFT.

In this paper, we compactify NRG data
for the two- and three-point Green's function
in the IR basis. Starting point is the representation of the  $n$-point time-ordered
correlation functions as a sum of  integrals convolving kernel functions and spectral densities. We further demonstrate that the IR basis cannot only be employed on the Matsubara axis but also on the real frequency axis.

The outline of the paper is as follows:
In Section~\ref{Sec_ImagGF=KerDen} the general representation of the
$n$-point Green's function in terms   of spectral densities and Kernel functions  is recapitulated, before turning to the specific spectral representation of the two- and three-point correlator which are at the focus of the present paper.
Section~\ref{Method} provides some information on the SIAM model and
NRG method.
In Section~\ref{sec:validation}, the NRG calculation of the two-point Green's function and  three-point Fermi-Bose vertex  is validated against the exact result in the atomic limit and  numerical quantum Monte Carlo (QMC) data for a finite hybridization, including an analysis of the error.
Section~\ref{IRbasis} briefly recapitulates the IR. 
In Section~\ref{realIRbasis} we present our key results, the IR representation of the NRG
two- and three-point correlator on the real axis.
Section~\ref{imagIRbasis} further demonstrates the  compactification, using real-frequency data as a starting point.
Finally, Section~\ref{Conclusion} provides a summary.

\section{Spectral density representation of two- and three-point fermionic correlators}\label{Sec_ImagGF=KerDen}

Let us presume the reader is familiar with the Matsubara and imaginary time formalism of quantum field theory  \cite{abrikosovmethods}
and define  the $n$-point fermionic imaginary-time Green's function
\begin{align}
\mathcal{G}(\tau_1,\ldots,\tau_n) = (-1)^{n-1}\mean{T_{\tau} A_1(\tau_1) \cdots A_n(\tau_n)},\label{N_IMAG_GF}
\end{align}
where the Wick operator $T_{\tau}$ orders the fermionic operators $A_i(\tau_i) = e^{H \tau_i} A_i e^{-H \tau_i}$ with $0 \leq \tau_i < \beta=1/T$ in descending chronology (with an additional minus sign for an odd permutation). For time-translational invariant systems, we can set $\tau_n=0$ as reference time and the Green's function reduces to
\begin{align}
&G(\tau_1,\ldots,\tau_{n-1}) = (-1)^{n-1} 
\mean{T_{\tau}\ \Biggl( \prod_{i=1}^{n-1} A_{i}(\tau_i) \Biggr)\ A_n}.\label{N-1_IMAG_GF}
\end{align}
We follow hereby the notation of Ref.\ \cite{kugler2021multipoint} using a calligraphic symbol in order to refer to the full $n$-point object, while a roman symbol for a correlator with $n-1$ independent arguments.
The time-ordered product can be formally expanded into a sum over permutations $S_{n-1} = \{\sigma_1, \sigma_2, \ldots, \sigma_{(n-1)!}\}$ of the time indices:
\begin{align}
&T_{\tau}\ \Biggl( \prod_{i=1}^{n-1} A_i(\tau_i) \Biggr) \nonumber\\
&\quad=\sum_{j=1}^{(n-1)!} \text{sgn}(\sigma_j)\Biggl( \prod_{i=1}^{n-2} \theta(\tau_{\sigma_j(i)}-\tau_{\sigma_j(i+1)}) \Biggr) \Biggl( \prod_{i=1}^{n-1} A_{\sigma_j(i)}(\tau_{\sigma_j(i)}) \Biggr), \label{eq:thetas}
\end{align}
where by $\sigma_j(i)$ we denote the place of the $i$-th index in the $j$-th permutation.
Note that all addends except for the chronologically ordered one are zero because of the Heaviside functions.
This expression can be further rearranged into kernel and spectral functions as follows (for a detailed derivation via the Lehmann representation see Appendix \ref{AppA})
\begin{align}
&G(\tau_1,\ldots,\tau_{n-1}) = \sum_{j=1}^{(n-1)!} \text{sgn}(\sigma_j) \int d^{n-1}\epsilon \nonumber\\
  &\qquad\times K_j(\tau_1,\ldots,\tau_{n-1};\epsilon_1,\ldots,\epsilon_{n-1}) \rho_j(\epsilon_1,\ldots,\epsilon_{n-1}),
  \label{eq:KernelGF}
\end{align}
since each permutation comes with a kernel function:
\begin{align}
&K_j(\tau_1,\ldots,\tau_{n-1};\epsilon_1,\ldots,\epsilon_{n-1}) \nonumber\\
&\quad= (-1)^{n-1} \Biggl( \prod_{i=1}^{n-2} \theta(\tau_{\sigma_j(i)}-\tau_{\sigma_j(i+1)}) e^{-\epsilon_i \tau_{\sigma_j(i)}} \Biggr)\ e^{-\epsilon_{n-1} \tau_{\sigma_j(n-1)}}, \label{GEN_IMAG_KERNEL}
\end{align}
convoluted with an associated spectral density:
\begin{align}
&\rho_j(\epsilon_1,\ldots,\epsilon_{n-1}) = \int d^{n}t\ \mean{\Biggl( \prod_{i=1}^{n-1} A_{\sigma_j(i)}(t_i)\ e^{\ii\epsilon_i t_i} \Biggr)\ A_{n}}.\label{GEN_REAL_RHO}
\end{align}
Here $\epsilon_{i}$ stands for excitation energies between two eigenstates (cf.\ Appendix \ref{AppA}), and the $d^{n-1}\epsilon$ and $dt^{n-1}$ integrations merely warrant $t_i=\tau_{\sigma_j(i)}$ reproducing the original Eqs.~\eqref{N-1_IMAG_GF}, \eqref{eq:thetas}.
This arrangement is very similar to Eq.\ (34a) ff.\  in Ref.\ \cite{kugler2021multipoint} except that we decompose here the time-translational invariant correlator and introduce an energy integral $d^{n-1} \epsilon$ to imply the connection to the spectral functions consisting of weights and excitations (similar to Eq.\ (28) in Ref.\ \cite{kugler2021multipoint}). Such a separation of the full Green's function into the product of kernels and spectral densities has been pointed out already in Ref.\ \cite{kobe1962spectral}.\\
The Green's function Eq.\ \eqref{N_IMAG_GF} and kernel Eq.\ \eqref{GEN_IMAG_KERNEL}, which consists of exponentially damped Heaviside step functions, are represented in Matsubara frequency notation using a Fourier transformation
\begin{align}
G(\nu_1,\ldots,\nu_{n-1}) = \int_{0}^{\beta} d^{n}\tau\ e^{\ii\sum_{i=1}^{n} \nu_i \tau_i}\ \mathcal{G}(\tau_1,\ldots,\tau_n).\label{N-1_IMAG_MF_GF}
\end{align}
The Matsubara Green's function in Eq.\ \eqref{N-1_IMAG_MF_GF} depends on $n-1$ frequencies, since the time-translational invariance in imaginary-time equals a energy conservation in the Matsubara representation. Within this paper we label fermionic Matsubara frequencies with $\nu = \frac{\pi}{\beta}(2k+1)$, respectively bosonic frequencies with  $\omega = \frac{\pi}{\beta}2k$, where $k \in \mathbb{Z}$ and $\beta = 1/T$.\\
For a deeper discussion on the analytic properties of the multipoint correlators we refer to Ref.\ \cite{shvaika2006spectral, shvaika2015spectral}.

\subsection{Fermionic spectral function of the two-point correlation function}

The spectral representation of the two-point Green's function is common textbook knowledge \cite{abrikosovmethods,mahanmany}. Let us nonetheless recapitulated it here for comparison and better understanding of the more complicated spectral representation of the $n$-point correlator derived below.
According to Eq.\ \eqref{N-1_IMAG_GF}, the two-point fermionic correlator describes the following thermal averaged dynamical amplitude
\begin{align}
G(\tau=\tau_1-\tau_2) = -\mean{T_{\tau}A(\tau)B(0)},\label{2_IMAG_GF}
\end{align}
where we substituted $A_1 = A$ and $A_2 = B$ in Eq.\ \eqref{N-1_IMAG_GF}. Following the former scheme and notation, the two-point correlator along the Matsubara frequency axis reads \cite{mahanmany}
\begin{align}
G(\ii\nu) = \int d\epsilon\ K_1(\ii\nu,\epsilon)\ \rho_1(\epsilon),\label{2_ANACONT}
\end{align}
with the following kernel function and spectral density:
\begin{align}
K_1(\ii\nu,\epsilon) &= \int_{0}^{\beta} d\tau\  e^{\ii\nu \tau}\ K_1(\tau,\epsilon) = \frac{e^{-\beta \epsilon}+1}{\ii\nu - \epsilon},\label{2_ANACONT_KER}\\
\rho_1(\epsilon) &= \int dt\ e^{\ii\epsilon t} \mean{A(t)B(0)}.\label{2_ANACONT_SD}\
\end{align}
For  $n=2$, the imaginary part of the associated retarded correlation function yields the kernel spectral density:
\begin{align}
\text{Im}[G(\ii\nu \rightarrow \nu+0^+)] &= \text{Im}[G^R(\nu)]\\
&= -\pi\ [e^{-\beta \nu}+1]\ \rho_1(\nu).\nonumber
\end{align}
which describes the linear response of a system to an external perturbation induced by operator B:
\begin{align}
G^R(\nu) &= \int dt\ e^{\ii\nu t}\ G^R(t)\label{2_RET_GF}\\
&= -\ii\int dt\ e^{\ii\nu t}\ \theta(t)\ \mean{[A(t),B(0)]_{+}}.\nonumber
\end{align}
Eq.\ \eqref{2_ANACONT} shows that the imaginary- and real-time representation of the fermionic two-point correlator is connected by the bare fermionic Kernel $K^F(\ii\nu,\omega)$. This yields a Fredholm integral equation \eqref{2_ANACONT}, which is rather difficult to solve and is a famous example for an inverse problem \cite{hansen2010discrete}.

\subsection{Spectral density representation of three-point correlator}

In this section we investigate the spectral representation for the three-point Fermi-Bose vertex with two fermionic $A,B$ and one bosonic $C$ operator:
\begin{align}
G(\tau_1,\tau_2) &= \mean{T_{\tau} A(\tau_1)B(\tau_2)C(0)}.\label{3_IMAG_GF}
\end{align}
This quantity forms the basis of several diagrammatic extensions of DMFT \cite{Ayral2016a,Stepanov2021,Krien2020,Krien2021}, and will be --besides the two-point correlator-- at the focus of our NRG analysis below.
There are two permutations, $\sigma_1=(12)$ and $\sigma_2=(21)$, for the time-ordering
and thus according to the last section we can represent the imaginary-time Fermion-Boson vertex  using two kernels and thermal spectral densities:
\begin{align}
G(\ii\nu_1,\ii\nu_2) =\sum_{j=1}^2 \int  d^{2}\epsilon\ K_j(\ii\nu_1,\ii\nu_2;\epsilon_1,\epsilon_2)\ \rho_{j}(\epsilon_1,\epsilon_2),\label{3_ANACONT}
\end{align}
where the kernel functions are again the Matsubara Fourier transformed counterparts to the exponentially damped Heaviside step functions as in Eq.\ \eqref{GEN_IMAG_KERNEL} for $n=3$ which stem from the time-ordering and time-evolution. Performing the according integrals for the above kernel functions leads to
\begin{align}
K_{1}(\ii\nu_1,\ii\nu_2;\epsilon_1,\epsilon_2)
=& g^{sg}(\epsilon_1,\epsilon_2)\delta_{\nu_1+\nu_2,0}K^{F}(\ii\nu_1,\epsilon_1)\\
&+ g_1^{FF}(\epsilon_1,\epsilon_2)K^{F}(\ii\nu_1,\epsilon_1)K^{F}(\ii\nu_2,\epsilon_2)\nonumber\\
&+ g_1^{FB}(\epsilon_1,\epsilon_2)K^{B}(\ii\nu_1+\ii\nu_2,\epsilon_1+\epsilon_2)\nonumber\\
& \times[1-\delta_{\nu_1+\nu_2,0}]K^{F}(\ii\nu_1,\epsilon_1)/(\epsilon_1+\epsilon_2),\nonumber
\end{align}
and
\begin{align}
K_{2}(\ii\nu_1,\ii\nu_2;\epsilon_1,\epsilon_2)
=& g^{sg}(\epsilon_1,\epsilon_2)\delta_{\nu_1+\nu_2,0}K^{F}(\ii\nu_1,\epsilon_2)\\
&+ g_2^{FF}(\epsilon_1,\epsilon_2)K^{F}(\ii\nu_1,\epsilon_1)K^{F}(\ii\nu_2,\epsilon_2)\nonumber\\
&+ g_2^{FB}(\epsilon_1,\epsilon_2)K^{B}(\ii\nu_1+\ii\nu_2,\epsilon_1+\epsilon_2)\nonumber\\
&\times [1-\delta_{\nu_1+\nu_2,0}]K^{F}(\ii\nu_2,\epsilon_1)/(\epsilon_1+\epsilon_2),\nonumber
\end{align}
where 
\begin{align}
g^{sg}(\epsilon_1,\epsilon_2) &= -\beta \delta_{\epsilon_1+\epsilon_2,0},\\
g_1^{FF}(\epsilon_1,\epsilon_2) &= e^{-\beta \epsilon_1}[1+e^{-\beta \epsilon_2}],\\
g_2^{FF}(\epsilon_1,\epsilon_2) &= -g_1^{FF}(\epsilon_2,\epsilon_1),\\
g_1^{FB}(\epsilon_1,\epsilon_2) &= [1-e^{-\beta (\epsilon_1+\epsilon_2)}][1-\delta_{\epsilon_1+\epsilon_2,0}],\\
g_2^{FB}(\epsilon_1,\epsilon_2) &= -g_1^{FB}(\epsilon_1,\epsilon_2).
\end{align}
The kernel functions $K_i$ can be represented as a linear combination of products of the bare fermionic [cf.\ Eq.\ \eqref{2_ANACONT_KER}] and bosonic kernels labeled with $F$ and $B$:
\begin{align}
K^{F}(\ii\nu_n,\epsilon) &= \frac{1}{\ii\nu_n - \epsilon},\\
K^{B}(i \omega_n,\epsilon) &= \frac{\epsilon}{i \omega_n - \epsilon}.
\end{align}
Each kernel function $K_i$ is convolved with an associated spectral density
\begin{align}
\rho_{1}(\epsilon_1,\epsilon_2) &= \int d^{2}t\ e^{i\epsilon_1 t_1} e^{i\epsilon_2 t_2} \mean{A(t_1)B(t_2)C(0)},\label{3_ANACONT_SD_1}\\
\rho_{2}(\epsilon_1,\epsilon_2) &= \int d^{2}t\ e^{i\epsilon_1 t_1} e^{i\epsilon_2 t_2} \mean{B(t_2)A(t_1)C(0)}.\label{3_ANACONT_SD_2}
\end{align}
in order to yield the Green's function. Finally, we emphasize that the three-point Green's function separates into a singular ($G_{sg}$) and normal ($G_{no}$) part, where the former scales with the inverse temperature:
\begin{align}
G(\ii\nu_1,\ii\nu_2) &= \delta_{\nu_1+\nu_2,0}\ \beta\ G_{sg}(\ii\nu_1,\ii\nu_2) + G_{no}(\ii\nu_1,\ii\nu_2)
\end{align}
where
\begin{align}
G_{sg}(\ii\nu_1,\ii\nu_2) =& -\!\int\! d^{2}\, \epsilon\ \delta_{\epsilon_1+\epsilon_2,0} K^{F}(\ii\nu_1,\epsilon_1) [\rho_{1}(\epsilon_1,\epsilon_2)\!+\!\rho_{2}(\epsilon_2,\epsilon_1)],\nonumber\\
G_{no}(\ii\nu_1,\ii\nu_2) =& \!\int\! d^{2}\, \epsilon\ K^{F}(\ii\nu_1,\epsilon_1)K^{F}(\ii\nu_2,\epsilon_2) \sum_{i=1,2}g^{FF}_{j}(\epsilon_1,\epsilon_2)\rho_{j}(\epsilon_1,\epsilon_2)\nonumber\\
+& \!\int\! d^{2}\, \epsilon\ [1-\delta_{\nu_1+\nu_2,0}] K^{B}(\ii\nu_1+\ii\nu_2,\epsilon_1+\epsilon_2)\nonumber\\
& \times \Big[ K^{F}(\ii\nu_1,\epsilon_1)g_1^{FB}(\epsilon_1,\epsilon_2)\rho_{1}(\epsilon_1,\epsilon_2)/(\epsilon_1+\epsilon_2)\nonumber\\
& + \ K^{F}(\ii\nu_2,\epsilon_1)g_2^{FB}(\epsilon_1,\epsilon_2)\rho_{2}(\epsilon_1,\epsilon_2)/(\epsilon_1+\epsilon_2) \Big].\nonumber
\end{align}
Let us note that we could have also defined thermal densities following the lines of Ref.\ \cite{shinaoka2018overcomplete}, given by $g^{\alpha}(\epsilon_1,\epsilon_2)\ \rho_i(\epsilon_1,\epsilon_2)$ , where $\alpha \in \{ sg,FF,FB\}$ and $i \in \{ 1,2 \}$. Within this notation the number of spectral densities increases as compared to Eq.\ \eqref{3_ANACONT}, while they are convolved with the bare fermionic and bosonic kernel functions from Eq.\ \eqref{2_ANACONT}.

\section{Model and Method}\label{Method}

We compute the spectral densities for the single-orbital impurity Anderson model (SIAM)
\begin{align}
H_{\text{SIAM}} &= \sum_{\sigma}\epsilon_d \cre{d}{\sigma}\ann{d}{\sigma} + U \cre{d}{\uparrow}\ann{d}{\uparrow} \cre{d}{\downarrow}\ann{d}{\downarrow} \label{SIAM_HAM}\\
&+ \sum_{k,\sigma} V_k [\cre{c}{k,\sigma}\ann{d}{\sigma} + h.c.] + \sum_{k,\sigma} \varepsilon_{k} \cre{c}{k,\sigma}\ann{c}{k,\sigma}\nonumber,
\end{align}
where the localized impurity electrons are represented by the second quantization operators $\{ \ann{d}{\sigma}, \cre{d}{\sigma} \}$ with spin $\sigma$ and interact with an effective Coulomb interaction $U$. The impurity electrons hybridize via $V_k$ to bath degrees of freedom $\{ \ann{c}{k,\sigma}, \cre{c}{k,\sigma} \}$. The non-interacting bath has a dispersion $\varepsilon_{k}$ such that the hybridization function is given by
\begin{align}
\Delta(\omega) = \sum_k |V_k|^2 \delta(\omega-\varepsilon_k). \label{HYB}
\end{align}
We solve the impurity model with a NRG routine \cite{wilson1975renormalization,krishna1980renormalization}. Let us here only briefly summarize the procedure and refer for more details to \cite{weichselbaum2012tensor,
bulla2008numerical}. As a first step the conduction band is divided up into logarithmic intervals $I^{+}_n = [\varepsilon^{+}_{n+1},\varepsilon^{+}_{n}]$ for $\omega > 0$, respectively $I^{-}_n = [\varepsilon^{-}_{n},\varepsilon^{-}_{n+1}]$ for $\omega < 0$,  using logarithmic discretization points $\varepsilon^{\pm}_{n} = \pm D \Lambda_{NRG}^{-n}$ with $n \in \mathbb{N}_0$ and $\Lambda_{NRG} > 1$ within the bandwidth $D$.

The conduction electron operators are then Fourier transformed within each logarithmic interval. It can be shown that thereby just the leading order term couples for a constant hybridization strength within each logarithmic interval to the impurity degrees of freedom. This approximation improves for decreasing $\Lambda_{NRG}$ and recovers the continuum in the limit $\Lambda_{NRG} \rightarrow 1$. Thus, the discretized hybridization function is represented by weights such that the total norm is conserved. Next, the thus discretized Hamiltonian is mapped by a unitary transformation onto a semi-infinite chain:
\begin{align}
H_{\text{WC}} &= \sum_{\sigma}\epsilon_d \cre{d}{\sigma}\ann{d}{\sigma} + U n_{\uparrow} n_{\downarrow} + V [\cre{f}{0,\sigma}\ann{d}{\sigma} + h.c.] \label{WILSON_HAM}\\
&+ \sum_{n=0,\sigma}^{\infty} t_n [\cre{f}{n+1,\sigma}\ann{f}{\sigma} + h.c.] + \sum_{n = 0,\sigma}^{\infty} g_{n} \cre{f}{n,\sigma}\ann{f}{n,\sigma}\nonumber.
\end{align}
This Hamiltonian describes fermions $\{ \ann{f}{n,\sigma}, \cre{f}{n,\sigma} \}$ that hop with amplitudes $t_n$ along the semi-infinite Wilson chain with a site-index $n$ and each with an onsite energy $g_n$, where the latter vanish for a symmetric hybridization function. The impurity just couples to the first site of the Wilson chain described by the following operator
\begin{align}
\ann{f}{0,\sigma} = \frac{1}{V}\sum_k V_k \ann{c}{k,\sigma}.
\end{align}
For more details we refer to Ref.\ \cite{bulla2008numerical}. In case the hopping amplitudes decay exponentially along the Wilson chain, the eigenstate space can be iteratively truncated, keeping only the $N_{keep}$ lowest eigenstates and  adding one additional bath site at a time. This tridiagonalization procedure is in general not necessary, but saves computational cost \cite{footnotezshift}.

\begin{figure}[tb!]
\centering
\includegraphics[width=0.35\textwidth]{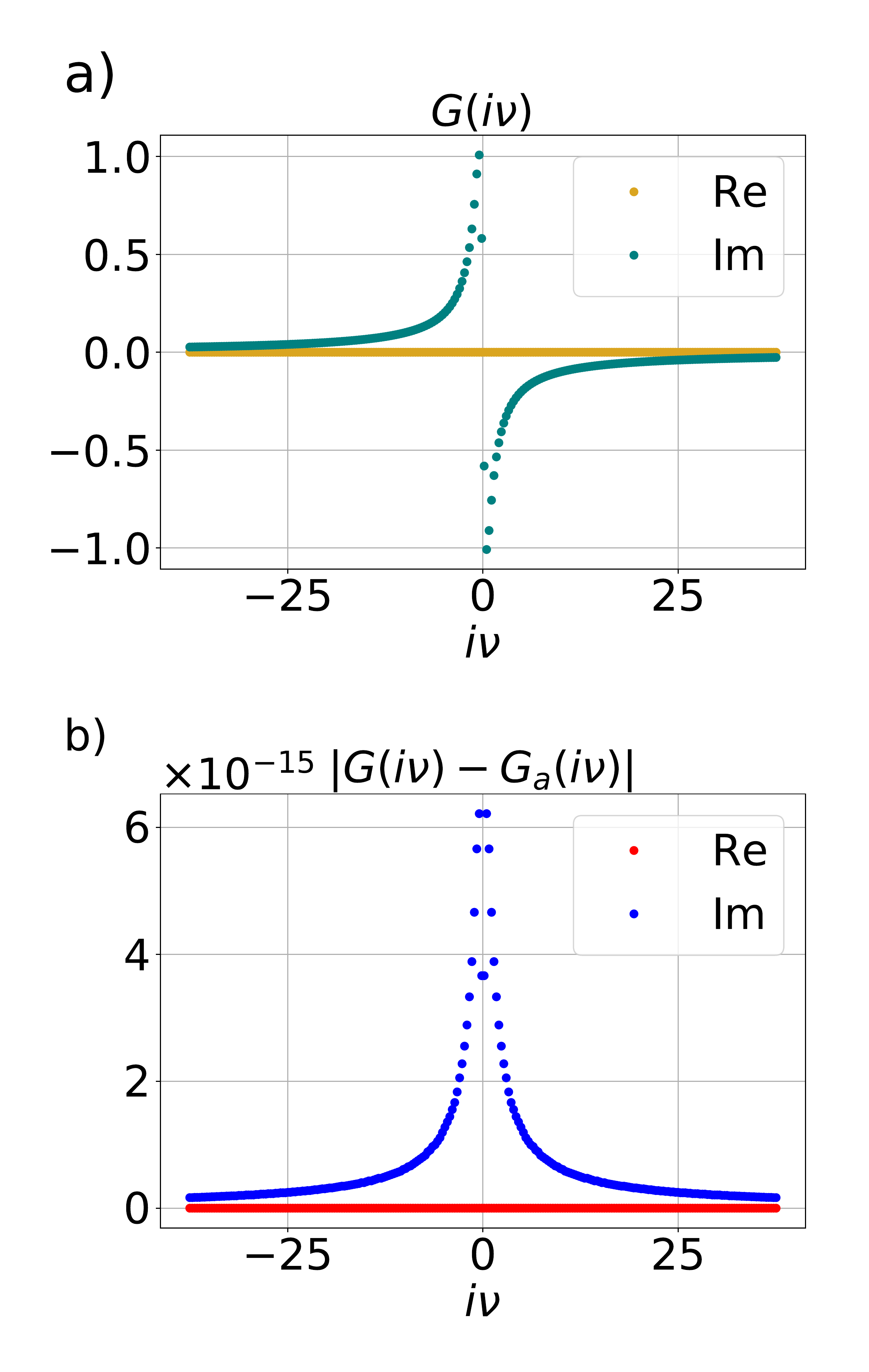}
\caption{Two-point Green's function in the atomic limit at an inverse temperature $\beta=20$, interaction strength $U=1.0$, and  half-filling. a)  $G(\ii\nu)$ as calculated by NRG with $N_{keep}=512$ and $\Lambda_{NRG}=2.0$. b)   Error  vs.\   the exact  $G_{a}(\ii\nu)$ [Eq.~\eqref{ATOMIC_LIMIT_GF}].}
\label{fig:Atom_2pt}
\end{figure}

The explicit computation of the spectral densities in Eq.\ \eqref{2_ANACONT_SD} and \eqref{3_ANACONT_SD_1}-\eqref{3_ANACONT_SD_2} are performed along the full-density matrix (FDM) formalism introduced in Ref.\ \cite{weichselbaum2007sum}. The FDM formalism provides a self-contained representation of the spectral densities with a high accuracy regarding the conservation of the spectral norm. However, it neglects processes connecting different shells within the Wilson chain. Further details on the implementation are discussed in the subsequent sections.

\section{Validation on the Matsubara axis} \label{sec:validation}
\subsection{Atomic limit}\label{Atom}

In order to analyze the performance of our NRG implementation, we start with the Hubbard atom, which corresponds to an isolated $s$ orbital with an effective Coulomb interaction $U$:
\begin{align}
H_{\text{ATOM}} &= \sum_{\sigma}\epsilon_d \cre{d}{\sigma}\ann{d}{\sigma} + U \cre{d}{\uparrow}\ann{d}{\uparrow} \cre{d}{\downarrow}\ann{d}{\downarrow}. \label{ATOM_HAM}
\end{align}
The numerical calculations are performed for an interaction strength $U=1.0$ and inverse temperature $\beta = 20$ with a local chemical potential $\epsilon_d = -U/2$ at half-filling.

\begin{figure*}[tb!]
\centering
\includegraphics[width=\textwidth]{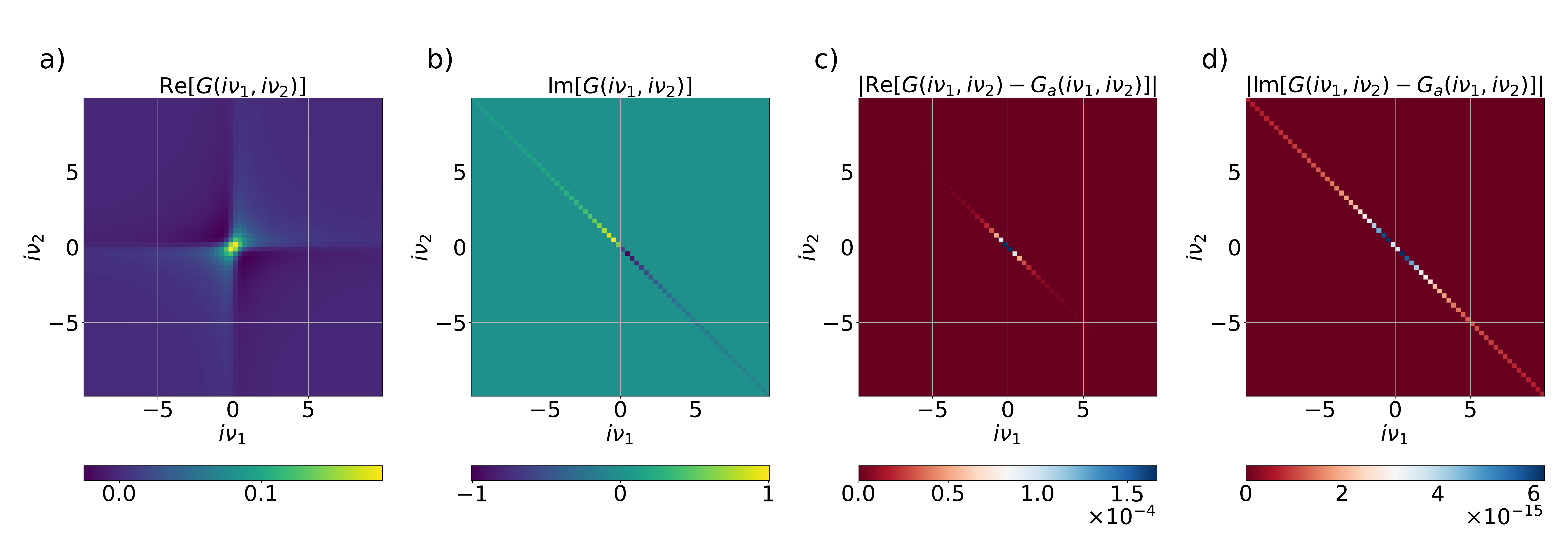}
\caption{Three-point Green's function in the atomic limit for $\beta=20$,  $U=1.0$, and half-filling: parts a) and b) show the real and imaginary part as calculated in NRG for $N_{keep}=512$ and $\Lambda_{NRG}=2.0$;  parts c) and d) the difference to the exact result $G_{a}(\ii\nu_1,\ii\nu_2)$ [Eq.~\eqref{3ptANA_RES} ff.].}
\label{fig:Atom_3pt}
\end{figure*}

\subsubsection{Two-point spectral density}
The imaginary-time spin-resolved Green's function reads
\begin{align}
G_a(\tau) = -\mean{T_{\tau}\ann{d}{\uparrow}(\tau)\ \cre{d}{\uparrow}(0)}.
\end{align}
In the atomic limit, the Anderson impurity can be solved analytically. Evaluating the spectral function yields 
\begin{align}
\mathcal{\rho}_{a}(\omega) &= \frac{1 + \exp[-\beta \epsilon_d]}{\mathcal{Z}} \delta(\omega-\epsilon_d)\\
&+ \frac{\exp[-\beta \epsilon_d] + \exp[-\beta (2\epsilon_d + U)]}{\mathcal{Z}} \delta(\omega-\epsilon_d-U)\nonumber
\end{align}
with the partition sum $\mathcal{Z} = \text{Tr}[e^{-\beta H}]$.

At half-filling and finite interaction strength the spectrum consists of two excitation peaks away from the Fermi level for the transitions from single to double and unoccupied impurity site, respectively. We convolve the kernels with the spectral functions to obtain the imaginary-time Green's function:
\begin{align}
G_{a}(\ii\nu) &= \int d\epsilon\ K^{F}(\ii\nu,\epsilon)\rho_{a}(\epsilon)\label{ATOMIC_LIMIT_GF} \\
&= \frac{1 + \exp[-\beta \epsilon_d]}{\mathcal{Z}} K^{F}(\ii\nu,\epsilon_d)\nonumber\\
&\phantom{=} + \frac{\exp[-\beta \epsilon_d] + \exp[-\beta (2\epsilon_d+U)]}{\mathcal{Z}} K^{F}(\ii\nu,\epsilon_d+U).\nonumber 
\end{align}
We plot in Fig.~\ref{fig:Atom_2pt}~a) the real and imaginary parts of the Green's function determined by NRG. Due to particle-hole symmetry the real part vanishes. As shown the imaginary part decreases for small Matsubara frequencies indicating that the original spectrum is insulating. In Fig.~\ref{fig:Atom_2pt}~b), we verify that analytic and numerical results are in excellent agreement.

\subsubsection{Three-point spectral densities}
Next we consider the local Fermi-Bose vertex
\begin{align}
G_a(\tau_1,\tau_2) = \mean{T_{\tau}\ann{d}{\uparrow}(\tau_1)\ \cre{d}{\uparrow}(\tau_2)\ n(0)}. \label{3ptANA_RES}
\end{align}
From Eq. \eqref{3_ANACONT_SD_1} and \eqref{3_ANACONT_SD_2} using a Lehmann representation we determine
\begin{align}
\rho_{a,1}(\epsilon_1,\epsilon_2) =& \frac{\exp[-\beta \epsilon_d]}{\mathcal{Z}}\delta(\epsilon_1-\epsilon_d-U)\delta(\epsilon_2+\epsilon_d+U),\\
\rho_{a,2}(\epsilon_1,\epsilon_2) =& \frac{\exp[-\beta \epsilon_d]}{\mathcal{Z}}\delta(\epsilon_1+\epsilon_d+U)\delta(\epsilon_2-\epsilon_d-U)\\
&+\frac{\exp[-\beta (2\epsilon_d+U)]}{\mathcal{Z}}\delta(\epsilon_1+\epsilon_d)\delta(\epsilon_2-\epsilon_d).\nonumber
\end{align}
Both spectral functions are centrosymmetric, i.e.\ $\rho_{a,i}(\epsilon_1,\epsilon_2) = \rho_{a,i}(-\epsilon_2,-\epsilon_1)$. As for the two-point correlator, we can separate the Fermi-Bose vertex
into  its normal and singular parts, i.e., $G_{a}=G_{a,no}+G_{a,sg}$ where
\begin{widetext}
\begin{align}
G_{a,no}(\ii\nu_1,\ii\nu_2) &= -\frac{1}{\mathcal{Z}}\left[\Big(1+\exp[-\beta\epsilon_d]\Big)\frac{1}{\ii\nu_1-\epsilon_d}\frac{1}{\ii\nu_2+\epsilon_d}+\Big(\exp[-\beta(2\epsilon_d+U)]+\exp[-\beta\epsilon_d]\Big)\frac{1}{\ii\nu_1-\epsilon_d-U}\frac{1}{\ii\nu_2+\epsilon_d+U}\right],\\
G_{a,sg}(\ii\nu_1,\ii\nu_2) &= \frac{1}{\mathcal{Z}}\left[\exp[-\beta \epsilon_d]\frac{1}{\ii\nu_1-\epsilon_d-U}+2\exp[-\beta (2\epsilon_d+U)]\frac{1}{\ii\nu_1-\epsilon_d-U}+\exp[-\beta \epsilon_d]\frac{1}{\ii\nu_1-\epsilon_d}\right].
\end{align}
\end{widetext}
Part a) and b) in Fig.\ \ref{fig:Atom_3pt} show real and imaginary part of the three-point Green's function against the fermionic Matsubara frequencies. For $\beta=20$ and $U=1$, the real part is dominated by the normal Green's function, whereas the imaginary part mostly depends on the singular contribution to the Green's function. At half-filling normal and singular Green's function satisfy
\begin{align}
G_{a,no}(\ii\nu_1,\ii\nu_2) &= G_{a,no}(\ii\nu_2,\ii\nu_1),\\
G_{a,sg}(\ii\nu_1,\ii\nu_2) &= -G_{a,sg}(\ii\nu_2,\ii\nu_1).
\end{align}
Comparison with the exact results, s.~Fig.~\ref{fig:Atom_3pt} c) and d), shows good agreement.

\subsection{Comparison with Quantum-Monte Carlo}\label{cfQMC}

After comparing our method to an exact solution for a simple model, we now turn to the full SIAM as defined in Eq.~\eqref{SIAM_HAM}. While no general analytic solution exists for the SIAM, quantum Monte Carlo (QMC) can, in principle, compute the result to arbitrary precision. In practice, however, the computational cost of the QMC algorithm scales unfavorably with decreasing temperature, which is not the case for the NRG procedure. 

\begin{figure}[tb!]
\centering
\includegraphics[width=0.35\textwidth]{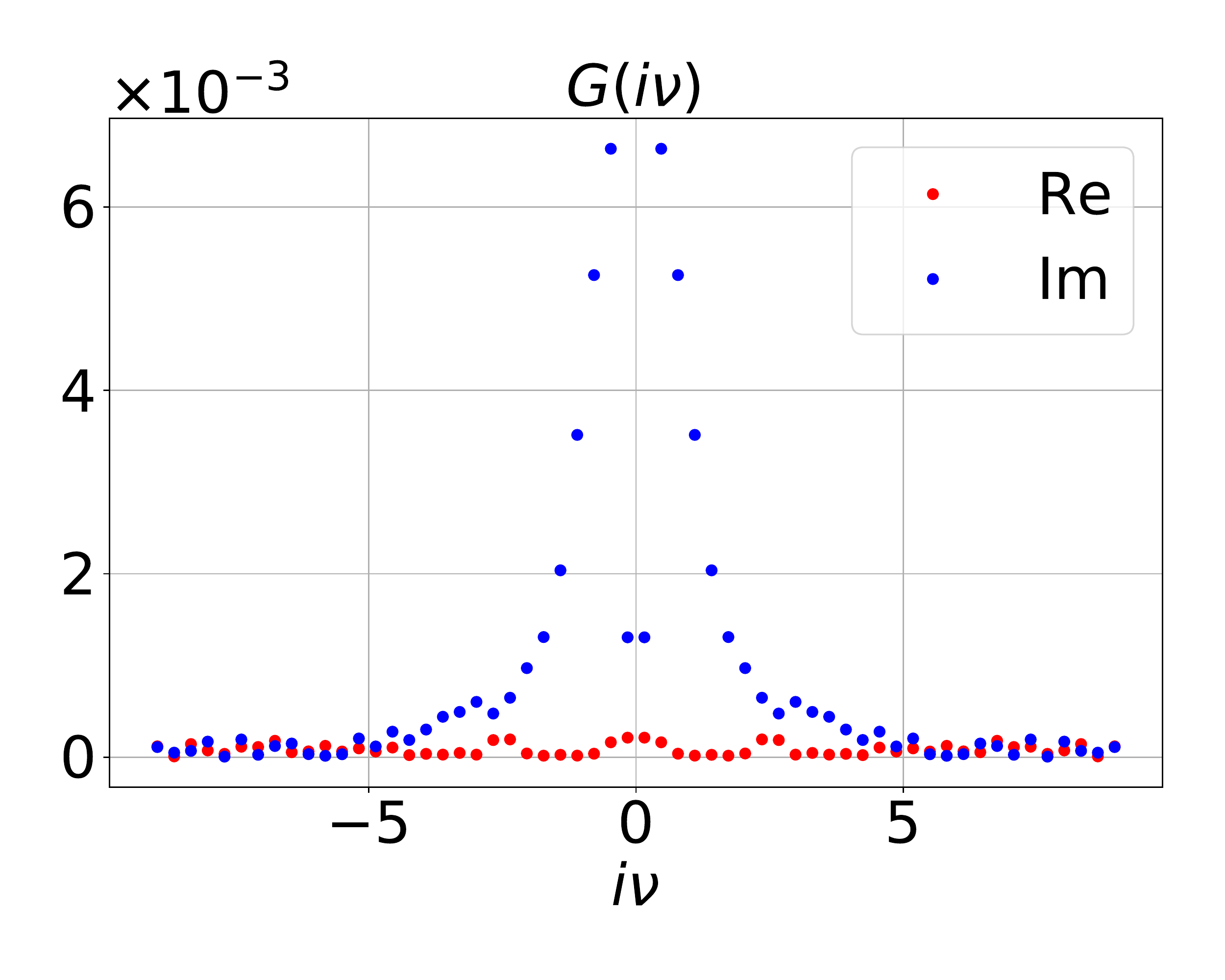}
\caption{Absolute difference of the two-point correlator $G(\ii\nu)$ between the NRG (with $N_{keep}=512$ and $\Lambda_{NRG}=2.0$) and QMC solution for the SIAM at $U=1.0$, $\beta=20$, constant hybridization $\Delta=0.1$ at half-filling. Real part (red) and imaginary part (blue). Statistical noise stems from QMC, while systematic deviations are from the NRG routine.}
\label{fig:CfQMC_2pt}
\end{figure}

Here we benchmark our NRG result to QMC data as generated by the \code{w2dynamics} \cite{Parragh2021,wallerberger2019} code, which uses continuous-time quantum Monte Carlo in the hybridization expansion \cite{Gull2011} (CT-HYB). We measured all quantities using the worm-sampling method \cite{Gunacker2015,Gunacker2016} and used order $10^8$ and $10^9$ measurements for the two-point and three-point correlator, respectively. 

For the AIM we choose a box shaped hybridization function [Eq.~\eqref{HYB}] $\Delta(\omega) = \Delta \Theta(D-|\omega|)$, where the band-width $D=1$ sets our unit of energy. We consider an interaction of $U=1\, D$  and the half-filled case, which corresponds to $\epsilon_d = -U/2$. The two-point and three-point correlators, as defined in Eq.~\eqref{2_ANACONT} and Eq.~\eqref{3_ANACONT}, are then measured at intermediate temperature $T=1/20\, D$. 

\begin{figure}[tb!]
\centering
\includegraphics[width=0.4\textwidth]{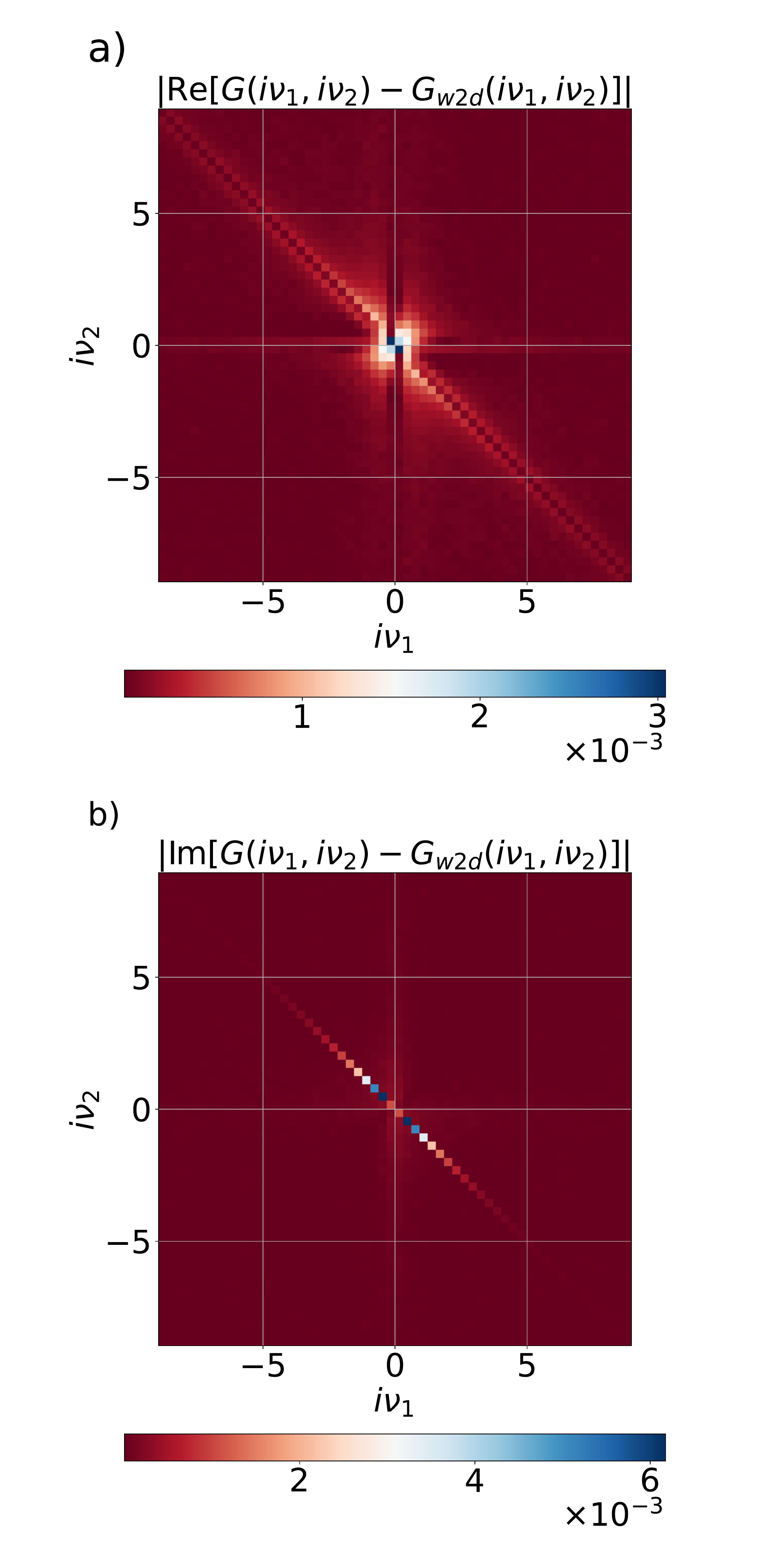}
\caption{Absolute difference of the three-point correlator $G(\ii\nu_1,\ii\nu_2)$ between the NRG (with $N_{keep}=512$ and $\Lambda_{NRG}=2.0$) and QMC solution for the SIAM at $U=1.0$, $\beta=20$, constant hybridization $\Delta=0.1$ at half-filling. a) Real part and b) imaginary part. Statistical noise stems from QMC, while systematic deviations are from the NRG routine.}
\label{fig:CfQMC_3t}
\end{figure}

Fig.~\ref{fig:CfQMC_2pt} and Fig.~\ref{fig:CfQMC_3t} display the absolute error between the QMC and NRG results for the two-point and three-point correlator, respectively. The background noise originates from the statistical error in the QMC data, while the structural deviation stems from NRG, which can mostly be observed in the low-frequency domain and along the anti-diagonal direction for the three-point function.

While QMC results can be improved by increasing the number of measurements, the NRG solution depends on the choice of $N_{keep}$ and $\Lambda_{NRG}$. Table~\ref{tab:cf} displays the total absolute error between QMC and NRG as a function of $N_{keep}$, which confirms the expected improvement of the NRG approximation. 




\begin{table}
\renewcommand{\arraystretch}{1.3}
\caption{Total absolute error between the NRG and QMC result for the two-point (2pt) and three-point (3pt) correlator  for the SIAM at $U=1$, $\beta=20$,  half-filling, various $N_{keep}$ of the NRG routine. For the error summation a box of $\pm 29$ frequencies for each argument was chosen, which corresponds to the frequency-windows displayed in Fig.~\ref{fig:CfQMC_2pt} and \ref{fig:CfQMC_3t}.}
\label{tab:cf}
\begin{center}
\begin{tabular}{|l|l|l|l|}
\hline
\multicolumn{1}{|l}{$|| \cdot ||_2$} & \multicolumn{1}{l|}{} & \multicolumn{2}{l|}{$\Lambda_{NRG}=2.0$}\\
\cline{3-4}
\multicolumn{2}{|l|}{[$10^{-2}$]} & 2pt & 3pt\\
\hline
\multirow{3}{*}{\rotatebox{90}{$N_{keep}$}} & 256 & 6.82 & 7.35\\ 
\cline{2-4}
 & 512 & 3.01 & 3.20\\
\cline{2-4}
 & 1024 & 1.38 & 1.61\\
\hline
\end{tabular}
\end{center}
\end{table}

\section{Intermediate representation}\label{IRbasis}

While it is difficult to find an effective discretization for the real frequency axis, due to the fact that
the spectral function can be arbitrarily ``peaky'', transitioning to the imaginary axis has a smoothening effect. This smoothing can be used to construct a rapidly converging representation.

We start by condensing Eqs.~(\ref{2_IMAG_GF}) and (\ref{2_RET_GF}) to:
\begin{align}
    G(\tau) &= \frac 1\pi \int d\epsilon\ \frac{e^{-\tau\epsilon}}{e^{-\beta\epsilon} \mp 1} \im G^R(\epsilon)\label{IR_KRAMERS_KRONIG} \\
    &= -\int d\epsilon\ \frac{e^{-\tau\epsilon}}{e^{-\beta\epsilon} \mp 1} \rho_1(\epsilon),\nonumber
\end{align}
where in the denominator we choose ``-'' for bosons and ``+'' for fermions. Frequently the kernel is represented by the dimensionless variables $x = \epsilon/\epsilon_{max} \in [-1,1]$ and $y = \tau/\beta-\tfrac{1}{2} \in [-1,1]$, with a characteristic   parameter $\Lambda_{IR}=\beta \epsilon_{max}$. The integral kernel in Eq.~(\ref{IR_KRAMERS_KRONIG}) admits a singular value expansion \cite{Hansen:SVE}:
\begin{equation}
    \frac{e^{-\tau\epsilon}}{e^{-\beta\epsilon} \mp 1} = \sum_{l=0}^\infty U_l^\pm(\tau)\ S_l^\pm\  V_l^\pm(\epsilon),
    \label{IR_SVE}
\end{equation}
where \{$S_l^\pm\}$ are the singular values in strictly decreasing order, $S_0 > S_1 > \ldots > 0$,
$\{U_l^\pm\}$ are the left singular functions, which form an orthonormal set
on the imaginary-time axis; and $\{V_l^\pm\}$ are the right singular functions
in  Eq.~\eqref{IR_SVE}, which form an orthonormal set on real frequencies. (For bosons, overlap on the real axis is
understood with respect to the measure $d\mu = \epsilon\,d\epsilon$.)

The singular functions $U_l^\pm$ and $V_l^\pm$ in Eq.~(\ref{IR_SVE}) are the so-called intermediate representation (IR) basis functions,
which can be used as representation for the imaginary-time Green's function~\cite{Shinaoka2017}:
\begin{equation}
G(\tau)=\sum_{l=0}^{L-1} U_l^\pm(\tau)\ G_l + E_L(\tau),\label{IR_GTAU_EXPAND}
\end{equation}
where $G_l = S_l\ \rho_l = S_l \int_\epsilon V_l(\epsilon)\rho(\epsilon)$ is a basis coefficient and $E_L$ is an error term associated
with truncating  the series after $L$ singular values.  As shown in \cite{Chikano:2018gd}, $S_l$ drops exponentially with $l$ and
grows only logarithmically with bandwidth in units of temperature~\cite{Chikano:2018gd}.  One also observes that the right singular functions $V_l$ are bounded, which implies the
truncated representation (\ref{IR_GTAU_EXPAND}) converges as
$\log E_L^{-1} = \bigO(L/\log(\beta\epsilon_{max}))$~\cite{Chikano:2018gd}.

In order to efficiently extract the imaginary-frequency basis coefficients $G_l$ from $G(\tau)$, we exploit the fact 
that the singular functions form a Chebyshev system~\cite{Wallerberger:2021}, similar in structure to orthogonal polynomials.
Hence, we choose a set of sampling points $\{\tau_1,\ldots,\tau_L\}$ as the roots of the highest-order basis function $U_L(\tau)$
and turn Eq.~(\ref{IR_GTAU_EXPAND}) into an ordinary least squares fit~\cite{Li:2020eu}:
\begin{equation}
    G_l = \arg\min_{G_l} \sum_{i=1}^L \bigg| G(\tau_i) - \sum_{l=0}^{L-1} U_l^\pm(\tau_i) G_l \bigg|^2.
    \label{IR_SPARSE}
\end{equation}
One empirically observes that the design matrix in Eq.~(\ref{IR_SPARSE}) is well-conditioned~\cite{Li:2020eu}, implying that the fitting error is consistent with the overall
truncation error $E_L$.

\section{Real-axis IR representation of spectral densities}\label{realIRbasis}
\subsection{Two-point correlator}

As already mentioned, the IR basis is compact on the imaginary-time axis, since there the expansion coefficients $G_l$ rigorously decay by virtue of the singular values $S_l$. Analytic expressions for the  real axis are not known and the signal can be arbitrarily more complicated. The aim of this paper is to numerically analyze
the prospects to use the IR coefficients also on the real axis.
The expansion coefficients of the spectral function projected onto the real-frequency singular basis functions read
\begin{equation}
    \rho_{j,l} = \int d\epsilon\ V_l(\epsilon)\ \rho_j(\epsilon),
    \label{IR_REAL}
\end{equation}
with $l=0,\ldots, L-1$ in order to provide a compact basis on the real axis for a given precision target 
\begin{equation}
  \epsilon_{error}=S_L/S_0 \label{eq:eerror}
\end{equation}
of the maximal ($L$'s) IR singular value (on the imaginary axis). Usually taking $L<L_{max}$ IR basis functions is sufficient thanks to their rapid decay on the imaginary axis. This means that real-frequency data which is  directly available from the imaginary axis, such as the density or the spectral weight in an interval $\epsilon \in [-1/\beta, 1/\beta]$, is retained, whereas other spectral features are averaged over.

On the real axis, the truncation of $\rho_l$ after $L$ IR coefficients acts as a "smearing" filter when back-transformed to real frequencies:
\begin{equation}
    \rho_{j,L}(\epsilon) := 
    \int d\epsilon'\  \left[ \sum_{l=0}^{L-1} V_l(\epsilon) V_l(\epsilon') \right] \rho_j(\epsilon').
\end{equation}
In order to get an intuition as to which features are smeared, we note that the right singular functions $\{V_l(\epsilon)\}$ also form a Chebyshev system~\cite{Karlin68}. Thus, there exists an associated Gauss quadrature rule on the real axis~\cite{Rokhlin:1996sve}. According to this, an integration over real frequency functions representable by a finite expansion in the basis set $\{V_l(\epsilon)\}$ equals a sum over the product of weights and the integrand evaluated at specific nodes. For ordinary polynomial basis functions, e.g. Legendre polynomials, the nodes are given by the roots of the highest polynomial needed to represent the function of interest and furthermore recursion relations for the weights exist. However, for a general Chebyshev system such as $\{V_l(\epsilon)\}$ the nodes and weights are a priori not known~\cite{Rokhlin:1996sve}.

\begin{figure}[tb]
\centering
\includegraphics[width=0.4\textwidth]{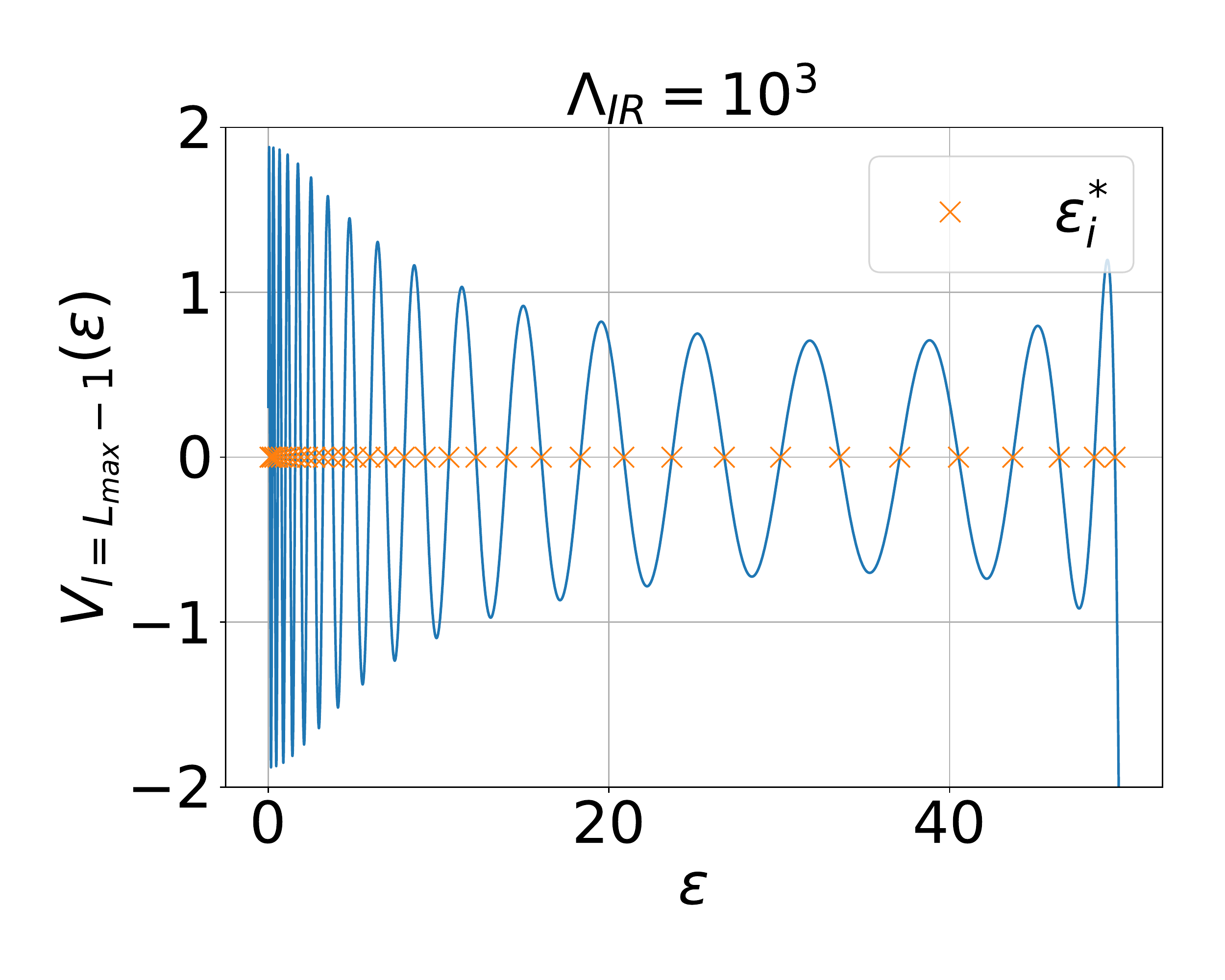}
\caption{Highest IR basis function $V_{L_{max}}(\epsilon)$ for $\Lambda_{IR}=10^3$. The density of its roots $\epsilon^*_l$ increases towards the Fermi energy. These roots are of special interest, since the IR basis function form a Chebyshev system (cf.\  main text).}
\label{fig:SING_FCT_VY}
\end{figure}

The density of the nodes of the highest computed IR basis function $V_{L_{max}}$ around some frequency $\epsilon$ is an indication of the resolution at that frequency. This means that smaller features in regions of more dense roots on the real axis have a more significant impact on the corresponding result on the imaginary axis. The zeros $\epsilon^*_i$ of $V_L$ around $\epsilon = 0$ are approximately distributed as follows (cf.\ Fig.\ \ref{fig:SING_FCT_VY}):
\begin{equation}
   \epsilon_i^* = o(C \exp(-\alpha |i - L/2|)),
   \label{IR_ZEROS_REAL}
\end{equation}
where $\alpha\to0.15(1)$ for $\Lambda_{IR}\to\infty$. The prefactor $C$ actually scales approximately with $1 / \Lambda_{IR}$ such that the density of roots close to the Fermi energy increase with smaller temperature and larger $\epsilon_{max}$. In conclusion, a truncated expansion in $\rho_{j,l}$ is expected to preserve features close to the Fermi edge and smear out features far away from it, on par with observations in numerical analytical continuation~\cite{Bryan90}.

\begin{figure*}[tb!]
\begin{minipage}{0.74\linewidth}
  \includegraphics[width=\linewidth]{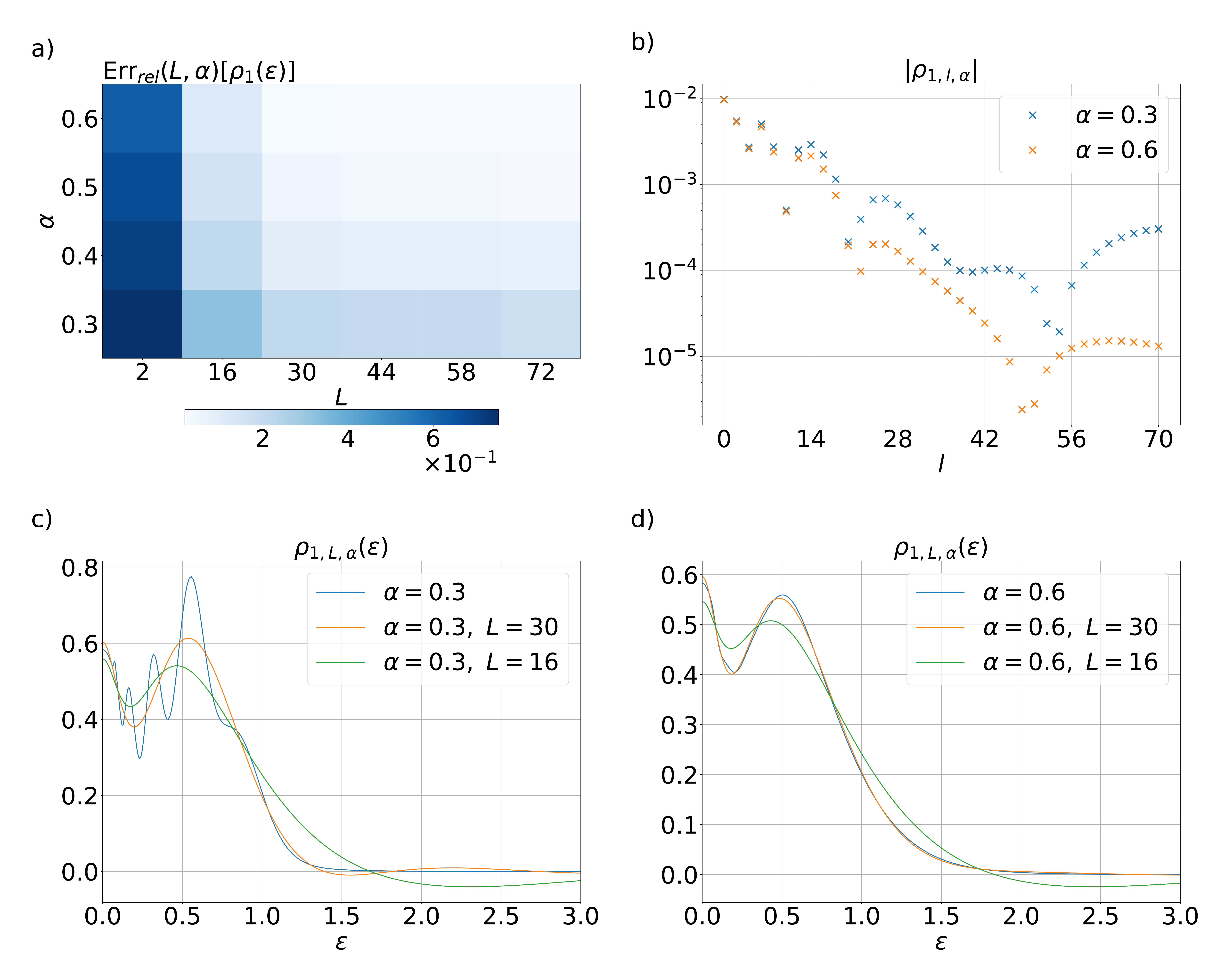}
\end{minipage}\hfill
\begin{minipage}{0.24\linewidth}
\caption{a) Relative error of the reconstructed two-point spectral density for the SIAM at  $\beta = 20$, $U=1.0$, constant hybridization $\Delta=0.1$, $N_{keep}=512$, $\Lambda_{NRG} = 2.0$ against the broadening parameter $\alpha$ and number of considered real-frequency IR basis coefficients $L$. With greater $\alpha$ and considering a larger IR basis space $L$, the error decreases. The further figures show additional information: b) Magnitude of expansion coefficient $\rho_{l,\alpha}$ over expansion coefficient $l$ for two different broadenings $\alpha$. The magnitude of the coefficients $\rho_{l,\alpha}$ shrinks with increasing  broadening parameter $\alpha$. c),d) We reconstruct the spectrum, keeping $L \in \{16,30\}$ IR basis functions and compare it to the original broadened NRG spectrum with $\alpha \in \{0.3,0.6\}$  (blue line).}
\label{fig:2pt_real_compress}
\end{minipage}\hfill

\begin{minipage}{0.74\linewidth}
\includegraphics[width=\linewidth]{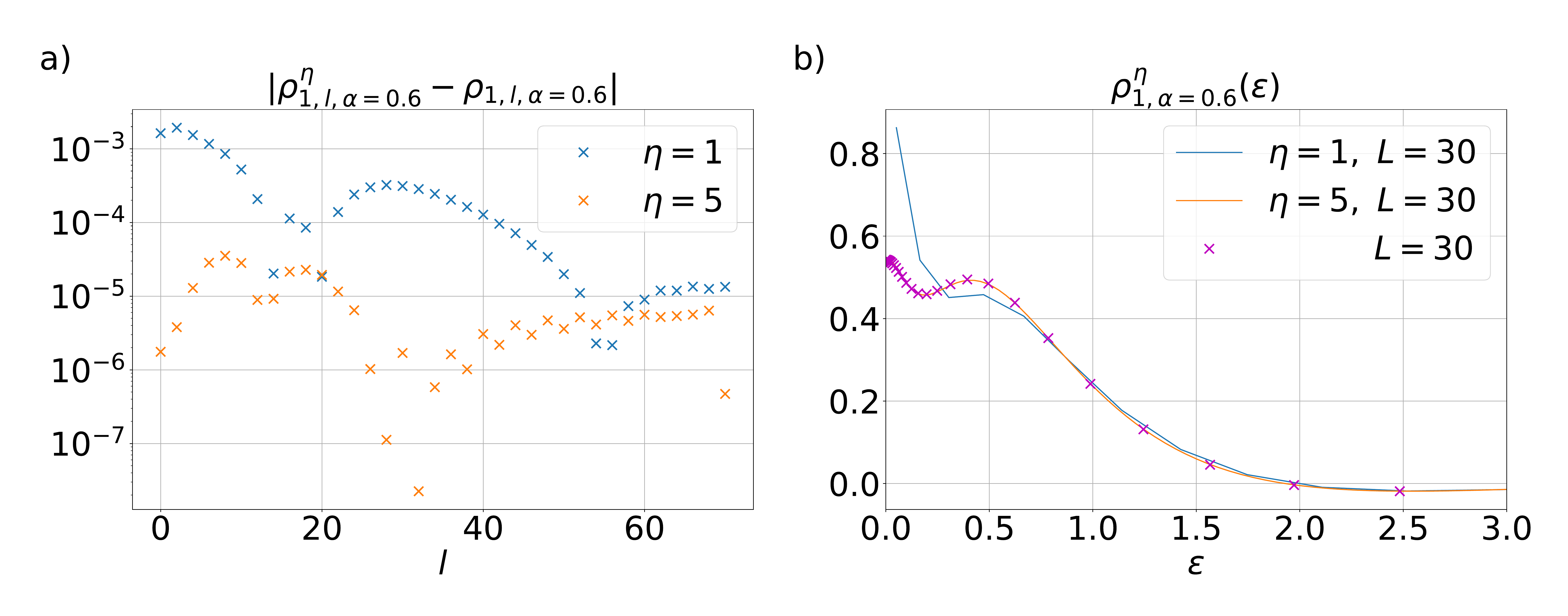}
\end{minipage}\hfill
\begin{minipage}{0.24\linewidth}
\caption{Thanks to the properties of the IR basis functions $V_l(\epsilon)$, only a small set of frequency points is needed to reproduce the original two-point function (same  parameters as in Fig.\ \ref{fig:2pt_real_compress}). a) In order to verify this, we show that the real-frequency IR basis coefficients for a fixed broadening $\alpha =0.6$ are well approximated for an oversampling factor $\eta=5$. b) This figure verifies that already $\eta=5$ is enough to yield the original signal along the real axis.}
\label{fig:2pt_real_compress_IR}
\end{minipage}
\end{figure*}

\begin{figure*}[tb!]
\begin{minipage}{0.71\linewidth}
\includegraphics[width=\linewidth]{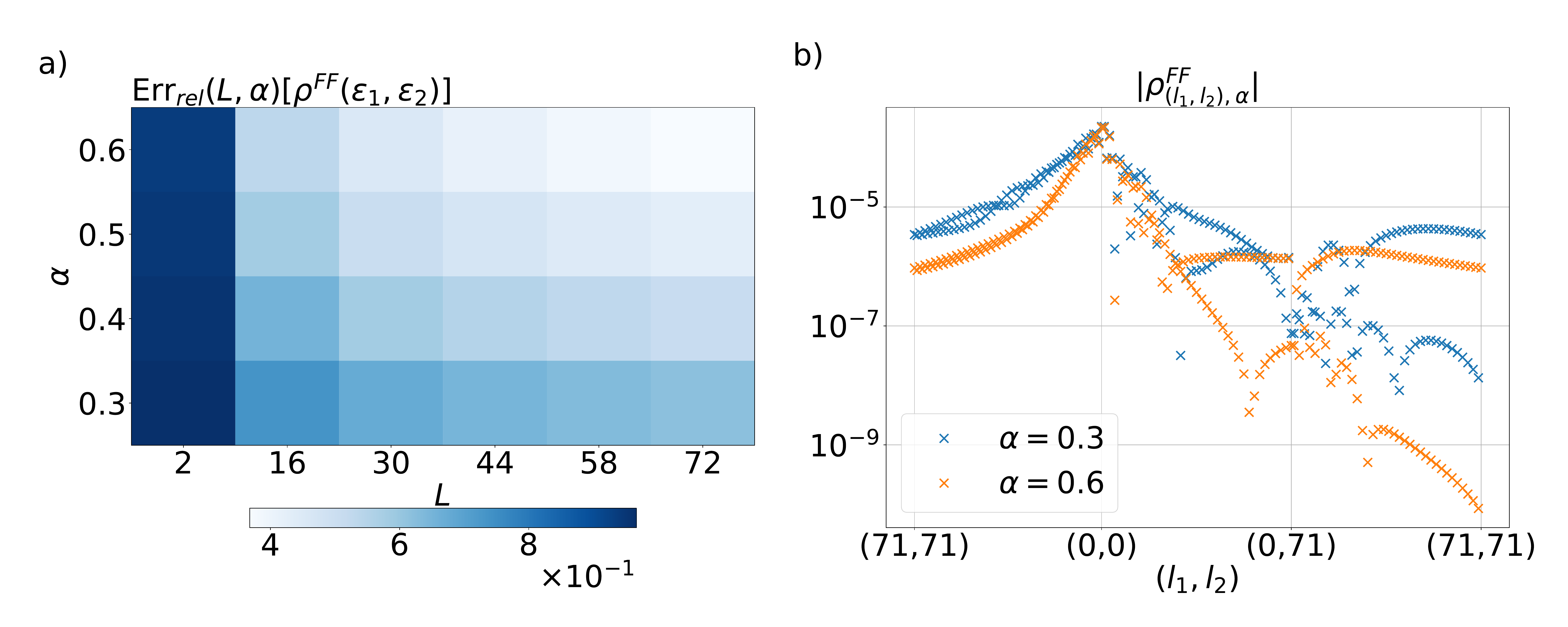}
\end{minipage}\hfill
\begin{minipage}{0.27\linewidth}
\caption{Relative error of the spectral density $\rho^{FF}_{12}(\epsilon_1,\epsilon_2)$ [cf. definition \eqref{SPECTRAL_DENSITY_12_FF}] of the three-point spectral densities for the SIAM at  $\beta = 20$, $U=1.0$, $\Delta=0.1$, half-filling, $N_{keep}=512$ and $\Lambda_{NRG}=2.0$. a) The discrete data is then broadened for a series of $\alpha$ and mapped onto the IR basis. The reconstructed signal, where we considered a finite number of IR basis coefficients $L$, is compared to the original spectral density by investigating the relative error. As discussed in the main text, the error reduces with greater broadening parameter $\alpha$. b) Real-frequency coefficients, which have the tendency to be smaller for larger $\alpha$.}
\label{fig:3pt_real_compress}
\end{minipage}
\end{figure*}

\begin{figure*}[tb!]
\includegraphics[width=\linewidth]{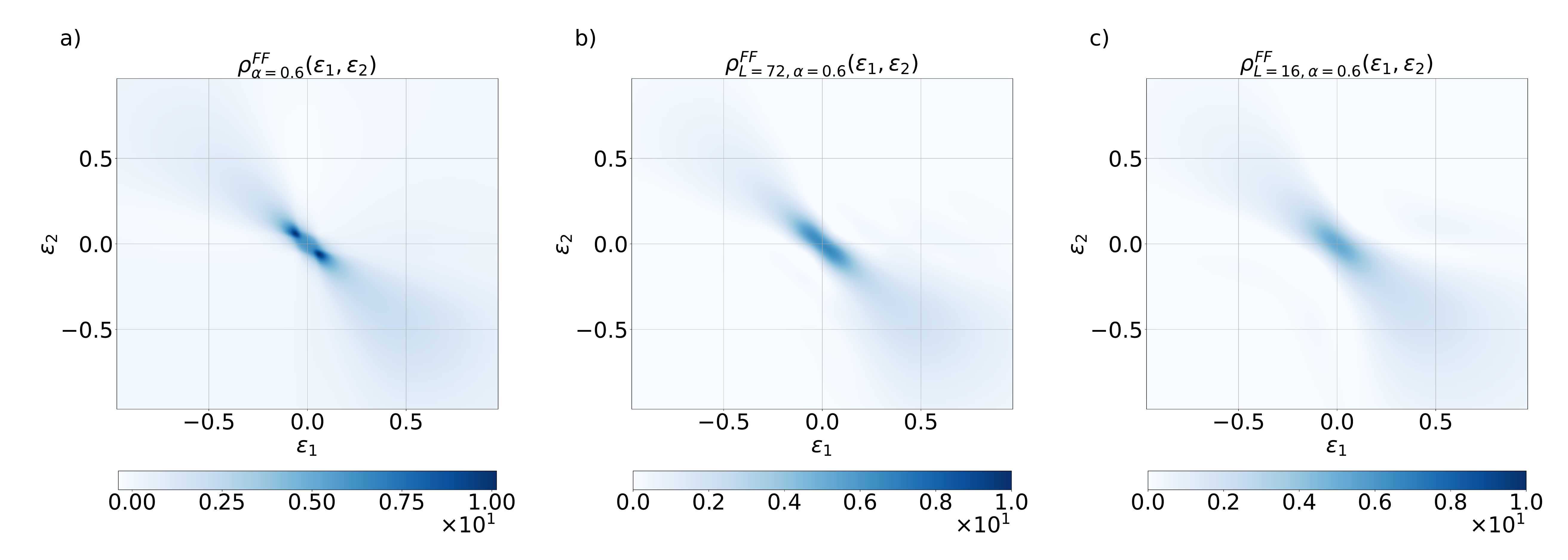}
\caption{a) Original three-pint spectrum $\rho^{FF}_{\alpha}$ after broadening the discrete NRG data with  $\alpha=0.6$ [cf.\ Eq.~\eqref{3ptbroadening}]. b) and c)  Reconstructed spectral density considering $L=72$ and $L=16$ IR basis coefficients, resepctively, as given by Eq.~\eqref{rho_FF_L72}. Close to the center $(\epsilon_1,\epsilon_2)=(0,0)$, we observe a smearing of finer details in the IR reconstructed spectral function.}
\label{fig:3pt_real_compress_IR_cf}
\end{figure*}

The IR basis has already been shown to be extremely useful as compression tool on the imaginary axis \cite{Chikano:2018gd}. In this paper we instead show how it can be applied to real-frequency objects. In the first step, we compute a discrete spectral function using our NRG routine. The peaky NRG spectrum is then broadened with a slightly modified version of the kernel proposed in Ref.\ \cite{weichselbaum2007sum}. The difference being an abrupt interpolation from Gaussian to log-Gaussian in our analysis:
\begin{align}
    \mathcal{K}_{\alpha}(\epsilon,\epsilon') = \label{BROAD_KERNEL}
    \begin{cases} 
      K^{G}(\epsilon,\epsilon') & |\epsilon'| < \epsilon_0, \\
      K^{GL}_{\alpha}(\epsilon,\epsilon') & |\epsilon'| \geq \epsilon_0,
   \end{cases}
\end{align}
with
\begin{align}
\mathcal{K}^{G}(\epsilon,\epsilon') &= \frac{1}{\sqrt{\pi \epsilon_0^2}} \exp{(-(\epsilon - \epsilon')^2/\epsilon_0^2)},\\
\mathcal{K}^{GL}_{\alpha}(\epsilon,\epsilon') &= \frac{\theta(\epsilon\epsilon')}{|\epsilon| \sqrt{\pi \alpha^2}} \exp\left[-\left(\frac1\alpha \log\left|\frac{\epsilon}{\epsilon'}\right| - \frac\alpha4\right)^2\right].
\end{align}
Here, the "smearing parameter" $\epsilon_0$ depends on microscopic details of the NRG flow and, according to Ref.\ \cite{weichselbaum2007sum}, is taken by a factor 2 smaller than the smallest energy scale in the system including also the Kondo temperature. The parameter $\alpha$ defines the broadening. Usually, in plain NRG calculations this parameter is set to $\alpha \sim 1/\sqrt{\Lambda_{NRG}}$, although it can be further reduced by using so-called z-shifts \cite{weichselbaum2007sum,footnotezshift}. In the following we analyze its impact on the compression. The broadened spectral function is then mapped onto the IR basis:
\begin{align}
    \rho_{j,l,\alpha} = \int d\epsilon\ \int d\omega\ \sum_{\omega_i} V_l(\epsilon)\ \mathcal{K}_{\alpha}(\epsilon,\omega_i)\rho_j(\omega),
    \label{RHOL_INT}
\end{align}
where on the right hand side of this equation we consider the discrete spectral density $\rho_j(\omega)$ of (weighted) delta peaks at the excitation energies.

As shown Fig.\ \ref{fig:2pt_real_compress}~b), the real-frequency IR coefficients tend to decrease stronger for a greater broadening parameter $\alpha$. To measure this on the frequency axis, we introduce the following norm and corresponding error:
\begin{align}
\text{Err}_{rel}(L,\alpha)[\rho_j(\epsilon)] = \frac{\underset{\epsilon}{\text{max}} |[\rho_{j,L,\alpha}(\epsilon)-\rho_{j,\alpha}(\epsilon)]|}{\underset{\epsilon}{\text{max}}|\rho_{j,\alpha}(\epsilon)|},
\end{align}
where $\rho_{j,L,\alpha}(\epsilon) = \sum_{l=0}^h{L-1} \rho_{j,l,\alpha} V_l(\epsilon)$ is again the back-transform from the IR basis to real frequencies (for the two-point case $j=1$). Later we also apply this measure to analyze the relative error of three-point functions. As expected the relative error shrinks with an increasing broadening parameter $\alpha$, see Fig.\ \ref{fig:2pt_real_compress}~a). Furthermore, the two lower panels show how the reconstructed signal improves with an increasing number of real-frequency IR basis coefficients. Already with a rather small broadening $\alpha \approx 0.6$, a finite number of IR basis coefficients $L \approx 30$ suffice to store most of the information of the original signal (cf.\ Fig.\ \ref{fig:2pt_real_compress}). Below an upper boundary $\alpha < 0.7$, we empirically observe, using a least-square fit, that
\begin{equation}
\text{Err}_{rel}(L,\alpha)[\rho_1(\epsilon)] \sim c_1(\alpha)\ L^{c_2(\alpha)},
\end{equation}
where $c_1(\alpha)$ is approximately a quadratic and $c_2(\alpha)$  a linear functions of $\alpha$. The fit has a root mean squared deviation of $\approx 0.14$ and the maximum standard error on the fitting parameters is reasonable for the data set shown in Fig.\ \ref{fig:2pt_real_compress}.

For fitting the IR coefficients directly to real frequency data, as obtained e.g.~from NRG, we rewrite the integral in Eq.\ \eqref{IR_REAL} as a minimization problem, where we replace the integral by a sum over a finite number of frequency points $\epsilon_i$:
\begin{align}
    \rho_{j,l,\alpha} &= \arg\min_{\rho_{j,l,\alpha}} 
    \sum_{i=1}^{N_{\epsilon}}\bigg|\rho_{j,\alpha}(\epsilon_i)-\sum_{l=0}^{L-1} V_{il}\ \rho_{j,l,\alpha} \bigg|^2 \label{RHOL_MIN}\\
    &= \sum_{i=1}^{N_{\epsilon}} V^{\oplus}_{li} \rho^{\phantom{x}}_{j,\alpha}(\epsilon_i),\nonumber
\end{align}
where $V^\oplus$ is the Moore--Penrose pseudoinverse of the matrix formed component-wise as 
$V_{il} = V_l(\epsilon_i)$~\cite{GolubVanLoan}.

For computing the pseudoinverse, we analyze the condition number of the matrix $V$, which is  minimized in case the set of frequencies $\epsilon_i$ includes a subset with roots of the highest singular value function $V_{L_{max}} (\epsilon)$. The number of roots  $\epsilon_i^*$ is of the order of the highest polynomial $L_{max}$, and, for a fixed temperature and bandwidth, the highest polynomial in the IR basis that needs to be computed  is given by $L_{max} \sim \log{\Lambda_{IR}} \log{\epsilon_{error}^{-1}}$ \cite{irbasis}, where $\epsilon_{error}$ [Eq.~\eqref{eq:eerror}] describes the upper bound for the truncation error for the singular value decomposition. Thus, the spectral function can be reconstructed up to an error $\text{Err}_{rel}(L,\alpha)$ on a dense frequency grid starting from its values at the roots of the highest polynomial $V_L(\epsilon)$:
\begin{align}
\rho_{j,\alpha}(\epsilon_i^{*}) = \int d\omega\ \mathcal{K}_{\alpha}(\epsilon_i^{*},\omega)\rho_j(\omega)\label{rho_root_alpha}
\end{align}
with the discrete NRG spectral density $\rho_j(\omega)$. The price for this reconstruction is a detour via the IR basis representation.
We further consider an oversampling factor $\eta$ according to which additional linearly separated points between each pair of adjacent roots, i.e.\ the total number of grid points used, yields approximately $N_{\epsilon} \approx L\ \eta$ frequencies. We mark this by adding a label to the broadened spectral density $\rho^{\eta}_{\alpha}(\epsilon)$, respectively the real-frequency IR basis coefficients $\rho^{\eta}_{l,\alpha}$. In case of $\eta =1$, the set of frequency points equals the roots $V_{L,\alpha}(\epsilon_i^{*})=0$.

In Fig.\ \ref{fig:2pt_real_compress_IR} a),
we compare the real-frequency IR basis coefficients
thus computed from the solution of Eq.\ \eqref{RHOL_MIN} for various oversampling factors to the integration of Eq.\ \eqref{RHOL_INT} on a very dense grid. It shows that with an increasing oversampling factor, the coefficients are closer to the result from the integral. On the real axis already $\eta = 5$ is sufficient to reproduce the original signal (s.\ Fig.\ \ref{fig:2pt_real_compress_IR} b)).

Of course, this approach entails a trade-off. The more peaky the spectral function is, i.e. the lower a broadening parameter $\alpha$ is used, the harder it can be compressed and thus more IR basis coefficients are needed. Consequently the number of roots at which  the spectral function in Eq.\ \eqref{rho_root_alpha} needs to be known increases. Second, we mention that for $\alpha > 1.0$ (outside the range of Fig.~\ref{fig:2pt_real_compress}) the results stagnate, i.e., the coefficients  $\rho_{j,l,\alpha}$ are not further decreasing with increasing $\alpha>0 \; \forall l$.

\subsection{Three-point spectral function}
We now turn to the three-point spectral function. As defined in Eq.\ \eqref{3_ANACONT}, we restrict our analysis to the contribution that comes with two fermionic kernels:
\begin{align}
    \rho^{FF}(\epsilon_1,\epsilon_2) = \sum_{j \in \{1,2\}} g^{FF}_{j}(\epsilon_1,\epsilon_2) \rho_j(\epsilon_1,\epsilon_2).
    \label{SPECTRAL_DENSITY_12_FF}
\end{align}
Again we consider the broadened spectral density and investigate its compressibility in the IR basis against broadening:
\begin{align}
    &\rho^{FF}_{\alpha}(\epsilon_1,\epsilon_2) \label{3ptbroadening}\\
    &= \int d^{2}\omega\ \mathcal{K}_{\alpha}(\epsilon_1,\omega_{1})\rho^{FF}(\omega_{1},\omega_{2})\mathcal{K}_{\alpha}(\omega_{2},\epsilon_2)\nonumber
\end{align}
with the discrete spectrum $\rho^{FF}(\omega_{1},\omega_{2})$. We emphasize that this broadening kernel is not physically motivated, i.e. we did not investigate if such a kernel leads, among other things, to the correct asymptotics. This is left for future work. Here, we are interested instead in the connection between broadening and compression in the IR basis of such a three-point real-frequency correlator.
Fig.\ \ref{fig:3pt_real_compress} b) shows that the real-frequency IR basis coefficients
\begin{align}
\rho^{FF}_{(l_1,l_2),\alpha} = \int d^{2}\epsilon\ \rho^{FF}_{\alpha}(\epsilon_1,\epsilon_2)V_{l_1}(\epsilon_1)V_{l_1}(\epsilon_2)
\end{align}
have the tendency to be smaller for $\alpha = 0.6$ than for $\alpha = 0.3$. We further demonstrate in Fig.\ \ref{fig:3pt_real_compress}  that for $L=72$, i.e.
\begin{align}
\rho^{FF}_{L=72,\alpha}(\epsilon_1,\epsilon_2) = \sum_{l_1,l_2 = 0}^{71} \rho^{FF}_{(l_1,l_2),\alpha} V_{l_1}(\epsilon_1) V_{l_2}(\epsilon_2),\label{rho_FF_L72}
\end{align}
the relative error of the resulting signal compared to the original spectral density reduces to $\text{Err}_{rel}(L,\alpha) \approx 0.4$ for $\alpha=0.6$.

\begin{figure*}[tb!]
\begin{minipage}{0.71\linewidth}
\includegraphics[width=\linewidth]{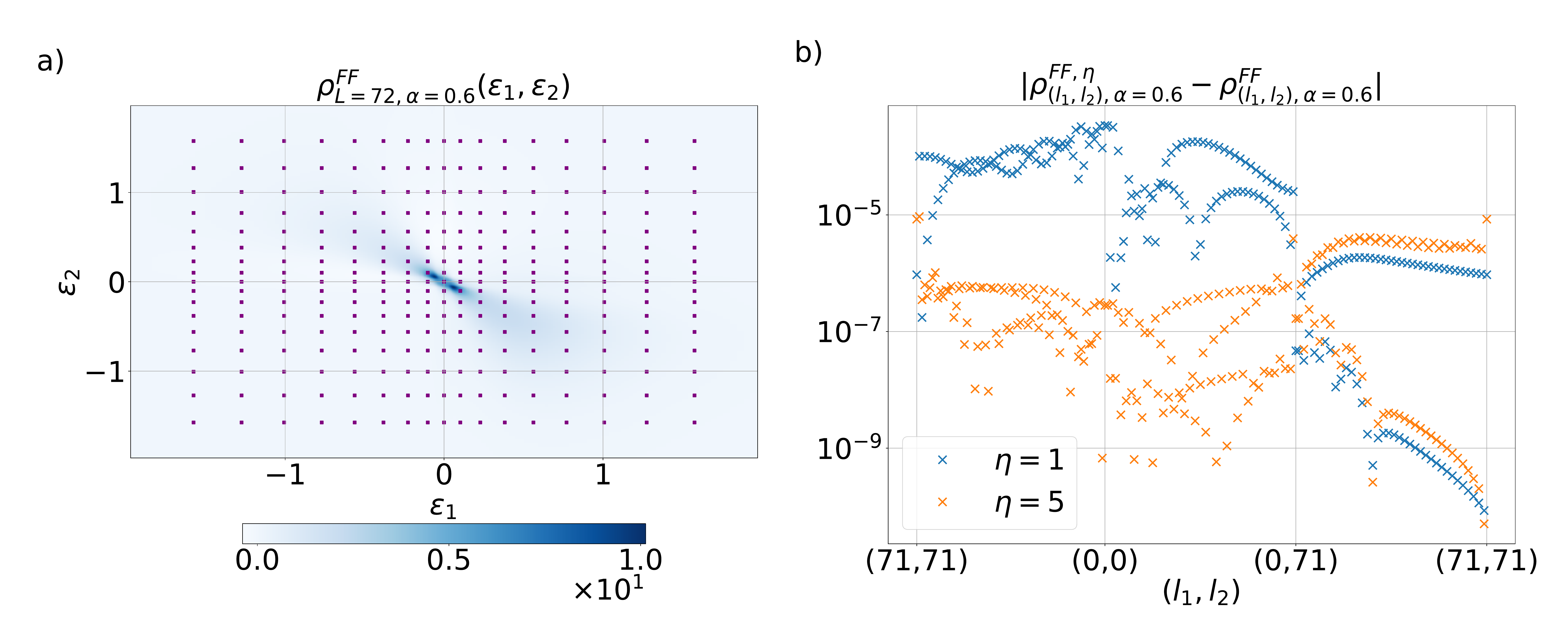}
\end{minipage}\hfill
\begin{minipage}{0.27\linewidth}
\caption{a) Spectral density $\rho^{FF}$ defined in Eq.~\eqref{SPECTRAL_DENSITY_12_FF} (false color plot; same parameters as in  Fig.\ \ref{fig:3pt_real_compress}) and roots of the highest polynomial $V_{L}(\epsilon_i^{*})=0$ (purple dots) at which the IR coefficients are computed. b) Comparison of  the real-frequency coefficients of the original signal [shown in panel a)] to the spectral density evaluated obtained by minimizing Eq.~\eqref{RHOL_MIN} at $N_\epsilon=L \eta$ roots $\epsilon_i^{*}$ for oversampling parameter $\eta=5$. The thus obtained  IR coefficients for the three-point vertex deviate only slightly,  maximally  of the order of $10^{-5}$.}
\label{fig:3pt_real_compress_IR}
\end{minipage}
\end{figure*}

In order to visualize the error in more detail and directly along the real frequency axis, we show in Fig.\ \ref{fig:3pt_real_compress_IR_cf} the original data $\rho^{FF}_{\alpha=0.6}$ next to the reconstructed signal as e.g.\ given by Eq.~\eqref{rho_FF_L72}. For $L=72$ the form of the spectral density is roughly reproduced in\ panel b). However details of the original data shown in panel a)  are smeared out, e.g. close to $(\epsilon_1,\epsilon_2)=(\pm 0.1,\mp 0.1)$. When reducing the number of considered IR basis coefficients to $L=16$ as shown in part c), this smearing becomes larger. According to this, the error between the original and reconstructed spectral densities reduces with increasing $L$. Altogether reproducing the three point vertex on the real axis with a reduced IR basis set appears to be more difficult than for the   two-point function in Fig.~\ref{fig:2pt_real_compress}. 
This might be because  the three-point vertex is either in general more complex compared to the two-point case, or because its asymptotics or its steep drop off sat small frequencies is more difficult to cover in the IR basis.

Following the approach for the two-point spectral function in Eq.~\eqref{RHOL_MIN}, we evaluate the spectral density just on the roots of the highest IR basis polynomial and consider an additional oversampling factor $\eta$ in order to add in-between additional frequency points.

\begin{figure*}[tb!]
\begin{minipage}{0.74\linewidth}
\includegraphics[width=\linewidth]{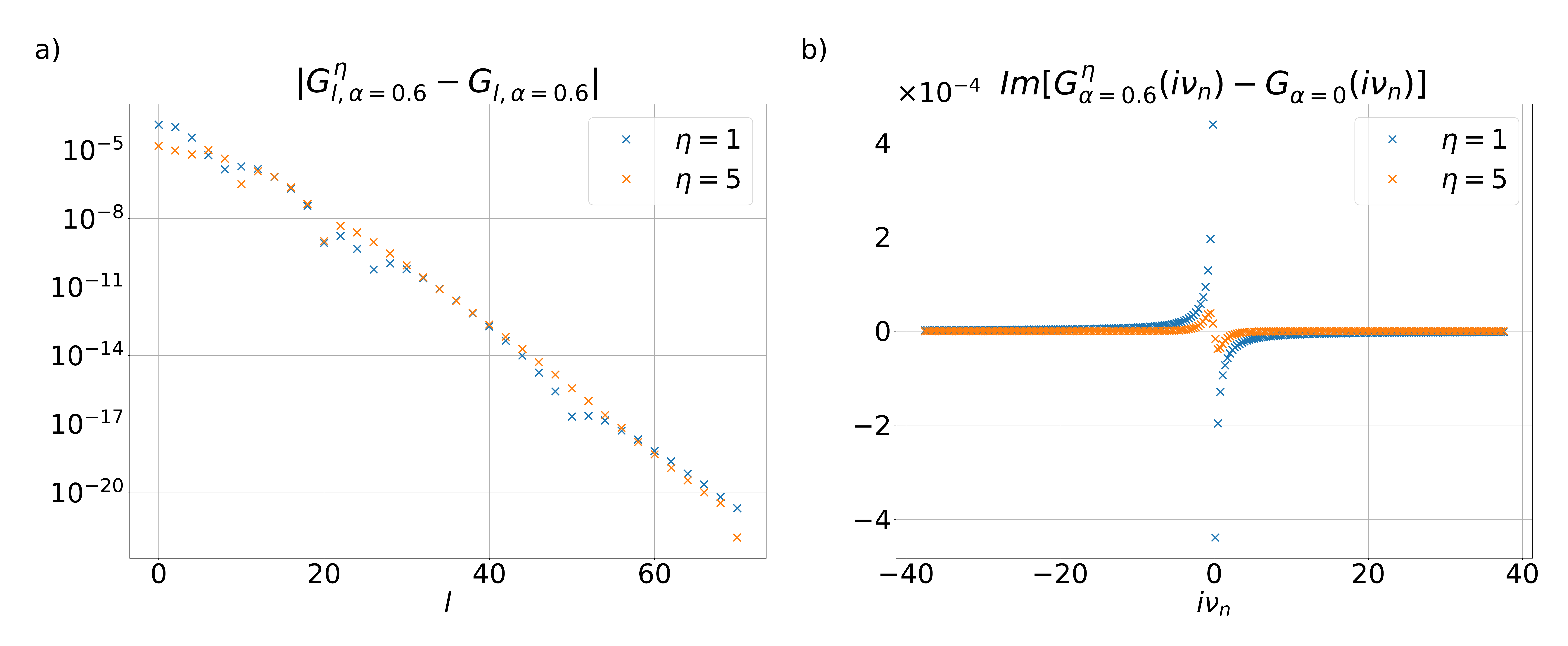}
\end{minipage}\hfill
\begin{minipage}{0.24\linewidth}
\caption{a) Error of the IR basis coefficients $G^{\eta}_{l,\alpha=0.6}$ compared to the exact solution for two oversampling factors $\eta \in \{1,5\}$ in comparison to the result from Eq.\ \eqref{RHOL_INT} (parameters as in Fig.\ \ref{fig:2pt_real_compress_IR}). The errors are rather similar. b) Absolute error of the two-point Matsubara frequency correlation function, which is of the order of $10^{-4}$ for $\eta=1$ and $10^{-5}$ for $\eta=5$.}
\label{fig:2pt_imag_compress_IR}
\end{minipage}
\end{figure*}

That is, we minimize for the three-point correlator
\begin{align}
    &\rho_{j,(l_1,l_2),\alpha} \label{RHOL_MIN_3pt}\\
    &= \arg\min_{\rho_{j,(l_1,l_2),\alpha}} \sum_{i_1,i_2=1}^{N_{\epsilon}}\bigg|\rho^{\phantom{x}}_{j,\alpha}(\epsilon_{i_1},\epsilon_{i_2})-\sum_{l_1,l_2=0}^{L-1} V^{\phantom{\oplus}}_{i_1 l_1} \rho^{\phantom{x}}_{j,(l_1,l_2),\alpha} V^{\oplus}_{l_2 i_2} \bigg|^2 \nonumber\\
    &= \sum_{i_1,i_2=1}^{N_{\epsilon}} V^{\oplus}_{l_1 i_1} \rho^{\phantom{x}}_{j,\alpha}(\epsilon_{i_1},\epsilon_{i_2}) V^{\phantom{\oplus}}_{l_2 i_2}\nonumber
\end{align}
using the definitions from Eq.\ \eqref{RHOL_MIN}.
The results are presented in Fig.\ \ref{fig:3pt_real_compress_IR}~a) shows the broadened spectral density, where we have marked the frequency points $(\epsilon_{i_1}^{*},\epsilon_{i_2}^{*})$ with $0 \leq i_1,i_2 \leq 71$ given by the roots of the highest order IR basis polynomial for $\Lambda_{IR} = 10^3$. In Fig.\ \ref{fig:3pt_real_compress_IR}~b) we see that the absolute error of the real-frequency IR basis coefficients compared to the original  value is reduced with $\eta$.

In summary, the IR basis on the real axis offers a possibility to compress the spectral density on the real axis. The rate of compression strongly depends on the smoothness of the data. For methods based on a diagonalization routine such as NRG, this is directly linked to the broadening kernel usually applied during the postprocessing routines. So far this is only an empirical observation, which needs to be further investigated.


\section{Compactification of imaginary-time correlation functions}\label{imagIRbasis}

Finally, we analyze the two- and three-point imaginary-time correlation function  using the IR basis representation. In the past the IR basis has been successfully used to compress data from diagrammatic or Monte Carlo approaches~\cite{NomotoPRB2020,NomotoPRL2020,NomuraRR2020,Isakov2020,Niklas2020,Pokhilko2021-uv,yeh2021electron,yeh2022relativistic,witt2021doping,Nagai:2019dea,nagai2021intrinsic,itou2021qcd,sakurai2021hybrid}. Compression of objects directly from the real-frequency axis has been unexplored previously in the literature. In this Section we cover how this compression reflects on the imaginary-time IR basis coefficients
and the Matsubara-frequency two-point Green's function and three-point Fermion-Bose vertex.

\subsection{Two-point correlator}
The computation of the real-frequency IR basis coefficients on the real axis for various oversampling factors has already been discussed in the former sections and is shown in Fig.\ \eqref{fig:2pt_real_compress_IR} for the two-point correlator.

\begin{figure*}[tb!]
\includegraphics[width=0.75\linewidth]{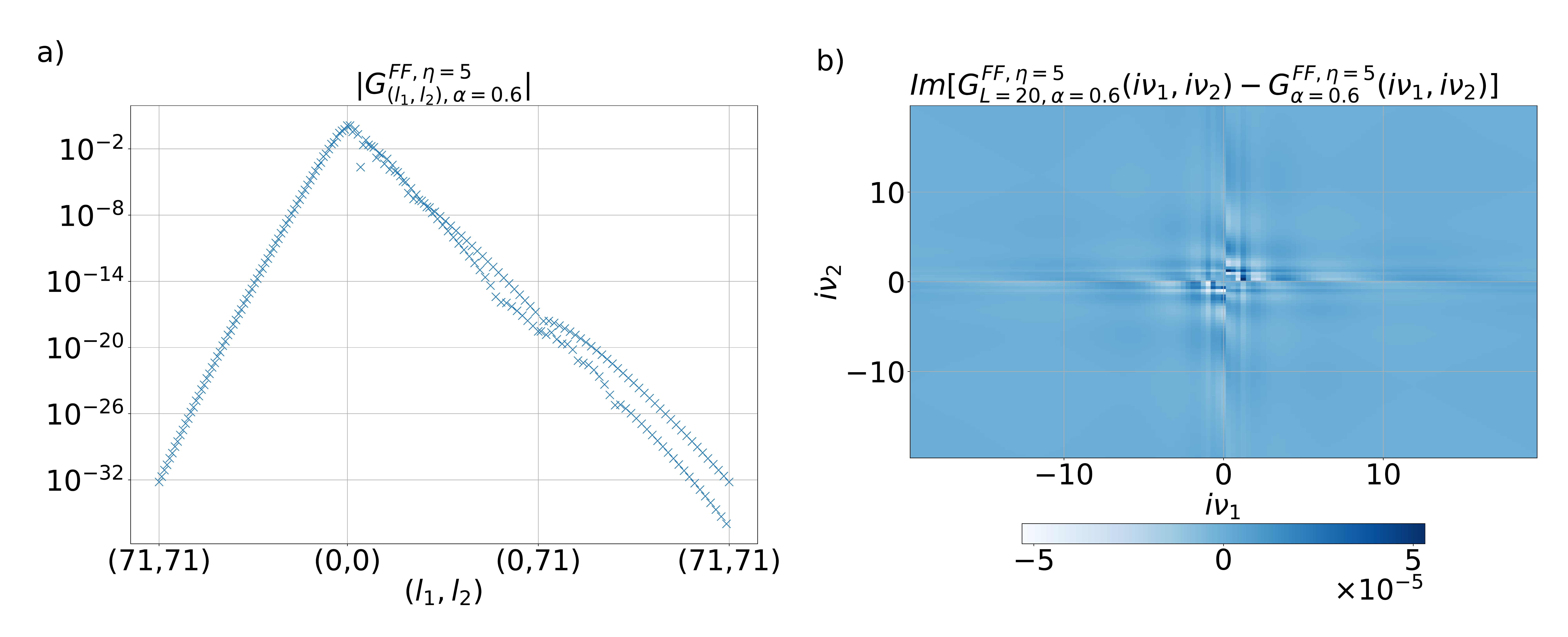}
\caption{a) IR basis coefficients $G^{FF,\eta=5}_{(l_1,l_2),\alpha=0.6}$ (parameters as in Fig.\ \ref{fig:3pt_real_compress}). As expected from the exponential decrease of the singular values, the coefficients are largest close to $l_1=l_2=0$. b) Additional error of the reconstructed three-point vertex
if we restrict the IR basis subspace  from $L=72$ to $0 \leq l_1,l_2 \leq L=20$. The absolute error is of the order of the corresponding IR coefficient $G^{FF,\eta=5}_{(l_1=20,l_2=L=20),\alpha=0.6}$ in panel a).}
\label{fig:3pt_imag_compress_IR}
\end{figure*}

\setlength{\belowcaptionskip}{-12pt}

We now multiply the coefficients  $\rho^{\eta}_{l,\alpha=0.6}$ elementwise with the singular values $S_l$ (cf.\ Eq.\ \eqref{IR_GTAU_EXPAND}) and display the result in Fig.\ \ref{fig:2pt_imag_compress_IR} a) again in comparison with the result from Eq.\ \eqref{RHOL_INT}. As expected the imaginary-frequency IR basis coefficients decay exponentially with increasing $l$ in Fig.\ \ref{fig:2pt_imag_compress_IR} a), and hardly deviate from their exact counterpart. Thus, the correlation function on the Matsubara axis is readily reconstructed by considering just a few imaginary-frequency IR basis coefficients. In Fig.\ \ref{fig:2pt_imag_compress_IR} b) we show further that since the coefficients  $G^{\eta}_{l,\alpha=0.6}$ for $\eta=1$ compared to $\eta=5$ differ on the order of $10^{-4}$ in  Fig.\ \ref{fig:2pt_imag_compress_IR} a), so do the two signals on the Matsubara frequency axis in  Fig.\ \ref{fig:2pt_imag_compress_IR} b).

\subsection{Three-point correlator}
Similarly, in Fig.\ \ref{fig:3pt_imag_compress_IR} a) the imaginary-frequency coefficients of the three-point correlation function 
\begin{align}
G^{FF,\eta=10}_{(l_1,l_2),\alpha=0.6}=S_{l_1l_2}\ \rho^{FF,\eta=10}_{(l_1,l_2),\alpha=0.6}
\end{align}
fall off very rapidly with their maximum at $l_1=l_2=0$.   Fig.\ \ref{fig:3pt_imag_compress_IR} b) shows the comparison of the reconstructed signal, when we consider just $L=20$ coefficients for both $l_1$ and $l_2$ in comparison with the full correlation function. Specifically, we visualize the error $Im[E_L(\ii\nu_1,\ii\nu_2)]$ (equivalently defined as in Eq.\ \eqref{IR_GTAU_EXPAND}, but for the three-point case and in Matsubara notation). The error is given by the magnitude of the coefficient $G^{FF,\eta=10}_{(l_1=20,l_2=20),\alpha=0.6}$, after which the IR basis is cut. Due to the structure of the correlation function , the error is largest for small Matsubara frequencies.\\
Figs.~\ref{fig:2pt_imag_compress_IR} and \ref{fig:3pt_imag_compress_IR} empirically demonstrate that the concept of the IR basis is useful to systematically save storage space and computing time. Already the evaluation of the two-point real-frequency spectral function just on the roots of the largest real-frequency IR basis polynomial, i.e.\ for $\eta=1$, is sufficient to represent the signal with high precision.

\section{Conclusion}\label{Conclusion}

We have demonstrated the prospects of using the IR basis for real-frequency NRG data. On the imaginary axis, the singular-value decomposition of the imaginary-time kernel functions  analytically guarantees an exponential convergence with the number of 
IR basis functions. For real times or frequencies such  a rigorous theorem does not exist.   We have, however, demonstrated empirically that a compactification to $L^2$ data points with $L=20 \ldots 30$ IR basis functions is possible for the real-frequency two-point correlator. Requirement for a fast convergence is broadening  the NRG delta-peaks. However only  a broadening factor $\alpha$ commonly employed in NRG  anyhow before $z$-shifts were introduced \cite{weichselbaum2007sum,footnotezshift} is needed.

In case of the three-point vertex it became obvious that the number of considered IR basis coefficients determines how well details of the spectral density are resolved. We have shown that for $L=72$ the very peaky structure at small frequencies  is smeared out. The  form of the spectral density is still reproduced. Overall the tendency of the error reduction with increasing IR basis coefficients $L$ holds.

Extrapolating the error, we expect to reconstruct the three-point vertex  rather well for $L \approx 150$. The alternative is storing the $n$-point correlators directly at the iteratively calculated $N$ NRG eigenfunctions \cite{lee2021computing}, requiring $N^{n-1}$ data. Usually in NRG,  $N=1000\ldots 10000$ is taken.\footnote{For four-point correlators evaluated in \cite{lee2021computing} this albeit had to be somewhat reduced because of memory constraints.} Using the IR basis results in a compactification by a factor of $(N/L)^{n-1}$, i.e., by the order of $10^{3(5)}$ for three(four)-point correlators.

\acknowledgements

We thank Friedrich Krien, Fabian Kugler, Seung-Sup Lee, Jan von Delft, and Clemens Watzenb\"ock for valuable discussions. This work has been  supported by the FWF (Austrian Science Funds) through project  P32044. Calculations have been done in part on the Vienna Scientific Cluster (VSC).

\appendix

\section{Kernel spectral density representation}\label{AppA}

We start from the definition of the imaginary-time Green's function in Eq.~\eqref{N_IMAG_GF}  and here explicitly derive the kernel and spectral density representation of  Eqs.\ \eqref{eq:KernelGF}-\eqref{GEN_REAL_RHO}:
\begin{widetext}
\begin{align}
G(\tau_1,\ldots,\tau_{n-1}) &= (-1)^{n+1} \mean{T_{\tau}\ \Biggl( \prod_{i=1}^{n-1} A_{i}(\tau_i) \Biggr)\ A_n} = \sum_{\sigma \in S_{n-1}} \text{sgn}(\sigma) (-1)^{n+1} \Biggl( \prod_{i=1}^{n-2} \theta(\tau_{\sigma(i)}-\tau_{\sigma(i+1)}) \Biggr) \mean{\Biggl( \prod_{i=1}^{n-1} A_{\sigma(i)}(\tau_{\sigma(i)}) \Biggr)\ A_n} \label{eq:SDderivation}\\
&= \sum_{\sigma \in S_{n-1}} \text{sgn}(\sigma) (-1)^{n+1} \Biggl( \prod_{i=1}^{n-2} \theta(\tau_{\sigma(i)}-\tau_{\sigma(i+1)}) \Biggr) \Biggl( \sum_{m_i} \frac{e^{-\beta E_{m_1}}}{\mathcal{Z}} \Biggl( \prod_{i=1}^{n-1} \bra{m_i} A_{\sigma(i)}(\tau_{\sigma(i)}) \ket{m_{i+1}} \Biggr) \bra{m_n} A_n \ket{m_1}  \Biggr)\nonumber\\
&= \sum_{\sigma \in S_{n-1}} \text{sgn}(\sigma) (-1)^{n+1} \Biggl( \prod_{i=1}^{n-2} \theta(\tau_{\sigma(i)}-\tau_{\sigma(i+1)}) \Biggr) \Biggl( \sum_{m_i} \frac{e^{-\beta E_{m_1}}}{\mathcal{Z}} \Biggl( \prod_{i=1}^{n-1} \bra{m_i} A_{\sigma(i)}e^{\tau_{\sigma(i)}(E_{m_i}-E_{m_{i+1}})} \ket{m_{i+1}} \Biggr) \bra{m_n} A_n \ket{m_1}  \Biggr)\nonumber\\
&= \sum_{\sigma \in S_{n-1}} \text{sgn}(\sigma) (-1)^{n+1} \Biggl( \prod_{i=1}^{n-2} \theta(\tau_{\sigma(i)}-\tau_{\sigma(i+1)}) \Biggr) \Biggl( \int_{\epsilon_i} \sum_{m_i} \frac{e^{-\beta E_{m_1}}}{\mathcal{Z}} \Biggl( \prod_{i=1}^{n-1} \bra{m_i} A_{\sigma(i)}e^{\tau_{\sigma(i)}\epsilon_i} \ket{m_{i+1}} \delta(\epsilon_i -(E_{m_i}-E_{m_{i+1}})) \Biggr) \bra{m_n} A_n \ket{m_1}  \Biggr)\nonumber\\
&= \sum_{\sigma \in S_{n-1}} \text{sgn}(\sigma) (-1)^{n+1} \Biggl( \prod_{i=1}^{n-2} \theta(\tau_{\sigma(i)}-\tau_{\sigma(i+1)}) \Biggr) \int_{\epsilon_i} \Biggl( \prod_{i=1}^{n-1} e^{\tau_{\sigma(i)}\epsilon_i} \Biggr)\ \rho_{\sigma}(\epsilon_1,\ldots,\epsilon_{n-1}) \nonumber\\
&= \sum_{\sigma \in S_{n-1}} \text{sgn}(\sigma) \int_{\epsilon_i} K_\sigma(\tau_1,\ldots,\tau_{n-1};\epsilon_1,\ldots,\epsilon_{n-1}) \rho_\sigma(\epsilon_1,\ldots,\epsilon_{n-1}),\nonumber
\end{align}
\end{widetext}
where $\{ \ket{m_i} , E_{m_i} \}$ label eigenstates and corresponding eigenenergies of the Hamiltonian, which we introduced in the second line of Eq.~\ref{eq:SDderivation}. We further introduced in the last two steps the generalized kernel and spectral density given by:
\begin{widetext}
\begin{align}
&K_{\sigma}(\tau_1,\ldots,\tau_{n-1};\epsilon_1,\ldots,\epsilon_{n-1})
= (-1)^{n+1} \Biggl( \prod_{i=1}^{n-2} \theta(\tau_{\sigma(i)}-\tau_{\sigma(i+1)}) e^{-\epsilon_i \tau_{\sigma(i)}} \Biggr)\ e^{-\epsilon_{n-1} \tau_{\sigma(n-1)}},\\
&\rho_\sigma(\epsilon_1,\ldots,\epsilon_{n-1}) = \sum_{m_i} \frac{e^{-\beta E_{m_1}}}{\mathcal{Z}} \Biggl( \prod_{i=1}^{n-1} \bra{m_i} A_{\sigma(i)} \ket{m_{i+1}} \delta(\epsilon_i -(E_{m_i}-E_{m_{i+1}})) \Biggr) \bra{m_n} A_n \ket{m_1} = \int_{t_i} \mean{\Biggl( \prod_{i=1}^{n-1} A_{\sigma(i)}(t_i)\ e^{i \epsilon_i t_i} \Biggr)\ A_{n}}.
\end{align}
\end{widetext}

\bibliography{hybNRG_FBVlit}

\end{document}